\documentclass[12pt]{article}

\usepackage{cite}
\usepackage{epsfig}
\usepackage{amsmath}

\def\beq{\begin{equation}}
\def\eeq{\end{equation}}
\def\beqar{\begin{eqnarray}}
\def\eeqar{\end{eqnarray}}
\def\barr#1{\begin{array}{#1}}
\def\earr{\end{array}}
\def\bfi{\begin{figure}}
\def\efi{\end{figure}}
\def\btab{\begin{table}}
\def\etab{\end{table}}
\def\bce{\begin{center}}
\def\ece{\end{center}}

\def\text{\textstyle}

\def\al{\alpha}

\def\De{\Delta}

\newcommand{\Oa}{\mathswitch{{\cal{O}}(\alpha)}}
\newcommand{\Oaa}{\mathswitch{{\cal{O}}(\alpha^2)}}

\def\mathswitchr#1{\relax\ifmmode{\mathrm{#1}}\else$\mathrm{#1}$\fi}

\newcommand{\PW}{\mathswitchr W}
\newcommand{\PZ}{\mathswitchr Z}

\newcommand{\PH}{\mathswitchr H}
\newcommand{\Pe}{\mathswitchr e}

\newcommand{\Pf}{\mathswitchr f}

\newcommand{\Pt}{\mathswitchr t}

\def\mathswitch#1{\relax\ifmmode#1\else$#1$\fi}

\newcommand{\MW}{\mathswitch {M_\PW}}

\newcommand{\MZ}{\mathswitch {M_\PZ}}
\newcommand{\MH}{\mathswitch {M_\PH}}

\newcommand{\Mt}{\mathswitch {m_\Pt}}
\newcommand{\scrs}{{}}
\newcommand{\sw}{\mathswitch {s_{\scrs\PW}}}

\newcommand{\cw}{\mathswitch {c_{\scrs\PW}}}

\newcommand{\GF}{\mathswitch {G_\mu}}

\newcommand{\mt}{\Mt}

\newcommand{\tsf}{\theta\kern-.20em_{\tilde{f}}}
\newcommand{\tsfp}{\theta\kern-.20em_{\tilde{f}\prime}}
\newcommand{\tsq}{\theta\kern-.15em_{\tilde{q}}}

\newcommand{\mf}{m_f}

\newcommand{\gsim}
{\;\raisebox{-.3em}{$\stackrel{\displaystyle >}{\sim}$}\;}

\newcommand{\alps}{\alpha_{\mathrm s}}

\newcommand{\VL}{\left( \begin{array}{c}}
\newcommand{\VR}{\end{array} \right)}
\newcommand{\ML}{\left( \begin{array}{cc}}
\newcommand{\MLd}{\left( \begin{array}{ccc}}
\newcommand{\MLv}{\left( \begin{array}{cccc}}
\newcommand{\MR}{\end{array} \right)}

\newcommand{\re}{\mbox {Re}\,}
\newcommand{\im}{\mbox {Im}\,}

\newcommand{\gev}{\,\, \mathrm{GeV}}

\newcommand{\BC}{\begin{center}}
\newcommand{\EC}{\end{center}}
\newcommand{\BE}{\begin{equation}}
\newcommand{\EE}{\end{equation}}
\newcommand{\BEA}{\begin{eqnarray}}
\newcommand{\BEAnn}{\begin{eqnarray*}}
\newcommand{\EEA}{\end{eqnarray}}
\newcommand{\EEAnn}{\end{eqnarray*}}

\newcommand{\id}{{\rm 1\kern-.12em
\rule{0.3pt}{1.5ex}\raisebox{0.0ex}{\rule{0.1em}{0.3pt}}}}

\newcommand{\SLASH}[2]{\makebox[#2ex][l]{$#1$}/}

\newcommand{\pslash}{\SLASH{p}{.2}}

\def\draftdate{\relax}
\def\mda{\relax}
\def\mua{\relax}
\def\mla{\relax}
\def\draft{
\def\thtystars{******************************}
\def\sixtystars{\thtystars\thtystars}
\typeout{}
\typeout{\sixtystars**}
\typeout{* Draft mode!
         For final version remove \protect\draft\space in source file
*}
\typeout{\sixtystars**}
\typeout{}
\def\draftdate{\today}
\def\mua{\marginpar[\boldmath\hfil$\uparrow$]%
                   {\boldmath$\uparrow$\hfil}%
                    \typeout{marginpar: $\uparrow$}\ignorespaces}
\def\mda{\marginpar[\boldmath\hfil$\downarrow$]%
                   {\boldmath$\downarrow$\hfil}%
                    \typeout{marginpar: $\downarrow$}\ignorespaces}
\def\mla{\marginpar[\boldmath\hfil$\rightarrow$]%
                   {\boldmath$\leftarrow $\hfil}%
                    \typeout{marginpar:
$\leftrightarrow$}\ignorespaces}
\def\Mua{\marginpar[\boldmath\hfil$\Uparrow$]%
                   {\boldmath$\Uparrow$\hfil}%
                    \typeout{marginpar: $\Uparrow$}\ignorespaces}
\def\Mda{\marginpar[\boldmath\hfil$\Downarrow$]%
                   {\boldmath$\Downarrow$\hfil}%
                    \typeout{marginpar: $\Downarrow$}\ignorespaces}
\def\Mla{\marginpar[\boldmath\hfil$\Rightarrow$]%
                   {\boldmath$\Leftarrow $\hfil}%
                    \typeout{marginpar:
$\Leftrightarrow$}\ignorespaces}
\overfullrule 5pt
\oddsidemargin -15mm
\marginparwidth 29mm
}

\newcommand{\SinEff}{\mathswitch {\sin^2\theta^{\mbox{\footnotesize lept}}_{\mbox{\footnotesize eff}}}}
\newcommand{\sineff}{\sin^2\theta^{\mbox{\scriptsize lept}}_{\mbox{\scriptsize eff}}}
\newcommand{\SinEfff}{$\sin^2\theta^{f}_{\mbox{\footnotesize eff}}$}
\newcommand{\sinefff}{\sin^2\theta^{f}_{\mbox{\scriptsize eff}}}

\newcommand{\mz}{\mathswitch {\overline{M}_\PZ}} 
\newcommand{\mw}{\mathswitch {\overline{M}_\PW}} 
\newcommand{\GZ}{\mathswitch {\Gamma_\PZ}} 
\newcommand{\gz}{\mathswitch {\overline{\Gamma}_\PZ}} 
\newcommand{\GW}{\mathswitch {\Gamma_\PW}} 

\newcommand{\li}{{\rm Li}}

\begin{document}
\thispagestyle{empty}

\def\thefootnote{\fnsymbol{footnote}}

\begin{flushright}
DESY 06-108\\
ZH--TH 17/06
\end{flushright}

\vspace{.5cm}

\begin{center}

{\Large\sc {\bf Electroweak two-loop corrections\\[.5em] to the
 effective weak mixing angle}}
\\[3.5em]
{\large\sc 
M.~Awramik$^{1}$%
\footnote{email: Malgorzata.Awramik@desy.de},
M.~Czakon$^{2}$%
\footnote{email: mczakon@yahoo.com}
and
A.~Freitas$^{3}$%
\footnote{email: afreitas@physik.unizh.ch}
}

\vspace*{1cm}

{\sl
$^1$ II. Institut f\"ur Theoretische Physik,
      Universit\"at Hamburg, \\ Luruper Chaussee 149, D-22761 Hamburg,
      Germany\\
{\rm and}\\
Institute of Nuclear Physics, Radzikowskiego 152,
      PL-31342 Krak\'ow, Poland
      
\vspace*{0.4cm}

$^2$ Institut f\"ur Theoretische Physik und Astrophysik, Universit\"at
W\"urzburg, Am Hubland, D-97074 W\"urzburg, Germany\\
{\rm and}\\
Institute of Physics, University of Silesia, Uniwersytecka 4,
  PL-40007 Katowice, Poland

\vspace*{0.4cm}

$^3$ Institut f\"ur Theoretische Physik,
	Universit\"at Z\"urich, \\ Winterthurerstrasse 190, CH-8057
	Z\"urich, Switzerland
}

\end{center}

\vspace*{0.5cm}

\begin{abstract}

Recently exact results for the complete electroweak two-loop contributions to
the effective weak mixing angle were published. This paper illustrates the
techniques used for this computation, in particular the methods for evaluating
the loop diagrams and the proper definition of $Z$-pole observables at
next-to-next-to-leading order. Numerical results are presented in terms of
simple parametrization formulae and compared in detail with a previous result
of an expansion up to next-to-leading order in the top-quark mass. Finally, an
estimate of the remaining theoretical uncertainties  from unknown higher-order
corrections is given.

\end{abstract}

\def\thefootnote{\arabic{footnote}}
\setcounter{page}{0}
\setcounter{footnote}{0}

\newpage


\section{Introduction}

One of the most important quantities for testing the Standard Model or its
extensions is the sine of the effective leptonic weak mixing angle \SinEff. In
the global fit of the Standard Model to all relevant electroweak data, the
effective leptonic weak mixing angle has a strong impact on indirect
constraints on $\MH$. It can be defined through the effective vector and
axial-vector couplings, $v_l$ and $a_l$, of the $Z$ boson to leptons ($l$) at
the $Z$ boson pole. Writing the $Z$ boson-lepton vertex as $\Gamma[Zl^+l^-] = 
i\; \overline{l} \gamma^\mu (v_l+a_l \gamma_5)l \; Z_\mu$, one obtains
\begin{equation}
  \label{eq:def}
  \sineff = \frac{1}{4}
  \left(1+\re\frac{v_l}{a_l}\right).
\end{equation}
Experimentally, \SinEff\ is 
derived from various asymmetries measured around the $Z$ boson peak at
$e^+ e^-$ colliders after subtraction of QED effects. It can also be determined
from asymmetries measured at center-of-mass energies away from the $Z$ pole,
requiring a theoretical extrapolation in order to match it to \SinEff\ on the $Z$
pole.
The current experimental accuracy, 
$\sineff = 0.23147 \pm 0.00017$ \cite{exp}, could be improved by an order of
magnitude at a future
high-luminosity linear collider running in a low-energy mode at the $Z$ boson
pole (Giga$Z$)~\cite{gigaz}. This offers the
prospect for highly sensitive tests of the electroweak
theory~\cite{gigaztests}, provided that the accuracy of the theoretical
prediction matches the experimental precision.

Typically, the theoretical prediction of \SinEff\ within the Standard Model is
given in terms of the following input parameters: the fine structure constant
$\alpha$, the Fermi constant $\GF$, the $Z$-boson mass $\MZ$ and the top-quark
mass $\mt$ (and other fermion masses whenever they are numerically relevant).
The $W$-boson mass $\MW$ is calculated from the Fermi constant, which is precisely
derived from the muon decay lifetime. As a consequence, the computation of
\SinEff\ involves two major parts: the radiative corrections to the relation
between  $\GF$ and $\MW$, and the corrections to the $Z$-lepton vertex form
factors. The latter can be incorporated into the quantity $\kappa =
1+\De\kappa$, defined in the on-shell scheme,
\begin{equation}
  \sineff = \left(1- \MW^2/\MZ^2 \right)  \left(1+
  \De\kappa \right),
  \label{eq:kappa}
\end{equation}
At tree-level,  $\De\kappa = 0$ and the sine of the effective mixing angle
is identical to the sine of the on-shell weak mixing angle $\sin^2 \theta_{\rm
W} \equiv \sw =  1-\MW^2/\MZ^2$. The quantity $\De\kappa$ is only weakly
sensitive to $\MW$.

For the computation of the $W$-boson mass, the complete electroweak two-loop
corrections, including partial higher-order
corrections, have been carried out in Ref.~\cite{muon,muon2,AC2,mw}.
In
this report, the calculation of the corresponding contributions for the form
factor $\Delta\kappa$ and combined predictions for \SinEff\ will be discussed.

The quantum corrections to \SinEff\ have been under extensive theoretical study over
the last two decades.
The one-loop result~\cite{sirlin,1loop} 
involves large fermionic contributions from 
the leading contribution to the $\rho$~parameter, $\De\rho$,
which is quadratically dependent on the top-quark mass $\mt$, resulting from 
the top-bottom mass splitting~\cite{velt}. The correction $\De\rho$ enters both
in the computation of $\MW$ from the Fermi constant (for a discussion see e.g.
Ref.~\cite{muon,muon2}), as well as into the vertex
correction factor $\De\kappa$,
\begin{equation}
1+\De\kappa^{(\al)} = 1+\frac{\cw^2}{\sw^2} \De\rho + 
\De \kappa_{\mathrm{rem}}(\MH),
\label{eq:delrol}
\end{equation}
with $\cw^2 = \MW^2/\MZ^2$, $\sw^2 = 1 - \MW^2/\MZ^2$.
The remainder part $\De \kappa_{\mathrm{rem}}$ contains 
in particular the dependence on the Higgs-boson mass, $\MH$.

Beyond the one-loop order, resummations of the leading one-loop contribution
$\De\rho$ have been derived~\cite{resum,hollikee}. They correctly take into
account the terms of the form  $(\De\rho)^2$ and $(\De\al\De\rho)$. Here
$\Delta\alpha$ is the shift in the fine structure constant due to light
fermions, $\De\al \propto \log \mf$, which enters through the corrections to
the relation between  $\GF$ and $\MW$, since $\De\kappa = \De\kappa(\MW)$ is a
function of $\MW$. These resummation results have been confirmed and extended
by an explicit calculation of the pure fermion-loop  corrections at $\Oaa$
(i.e.\ contributions containing two fermion loops)~\cite{floops}. Recently, the
leading three-loop contributions to the $\rho$ parameter of ${\cal O}(\GF^3
\mt^6)$ and ${\cal O}(\GF^2 \alps \mt^4)$ for large top-quark mass \cite{mt6},
as well as ${\cal O}(\GF^3 \MH^4)$ for large Higgs mass 
\cite{radja} have been computed.

Higher order QCD corrections to \SinEff\ have been calculated at ${\cal O}(\al
\alps)$~\cite{qcd2} and for the top-bottom contributions at ${\cal O}(\al
\alps^2)$~\cite{qcd3} and ${\cal O}(\al
\alps^3)$~\cite{qcd4}. The ${\cal O}(\al \alps^2)$ contributions with light
quarks in the loops can be derived from eqs.~(29)--(31) in
\cite{qcd3light} and turn out to be completely negligible. 
For the electroweak two-loop contributions, only partial results using large
mass expansions in the Higgs mass \cite{ewmh2} and top-quark
mass~\cite{ewmtmh,ewmt4,ewmt2} have been known previously.
Concerning the expansion in $\mt$, the
formally leading term of ${\cal O}(\GF^2 \Mt^4)$~\cite{ewmtmh,ewmt4} and the
next-to-leading term of ${\cal O}(\GF^2 \Mt^2 \MZ^2)$~\cite{ewmt2} were found
to be numerically significant and of similar magnitude. Therefore, a
complete calculation of electroweak two-loop corrections to \SinEff\ beyond the
leading terms of expansions is desirable.

As a first step in this direction, exact results have been obtained for the
Higgs-mass dependence (i.e.\ the quantity  $\sin^2\theta^{\mbox{\footnotesize
lept}}_{\mbox{\footnotesize eff,sub}}(\MH) \equiv \sineff(\MH) - \sineff(\MH =
65 \gev))$ of the two-loop  corrections with at least one closed fermion loop
to the precision observables~\cite{floops,ewmhdep}.  They were shown to agree
well with the previous results of the top-quark mass expansion~\cite{gsw}.

\begin{figure}[tb]
\begin{center}
\raisebox{3cm}{(a)}\psfig{figure=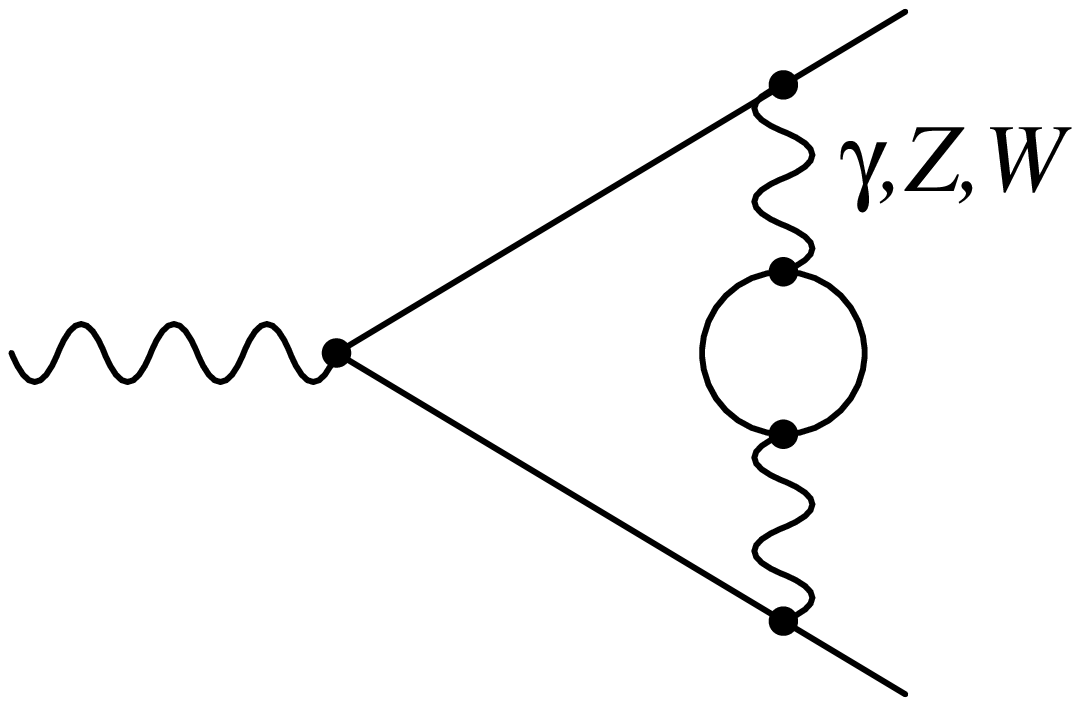, width=5cm} 
\raisebox{3cm}{(b)}\psfig{figure=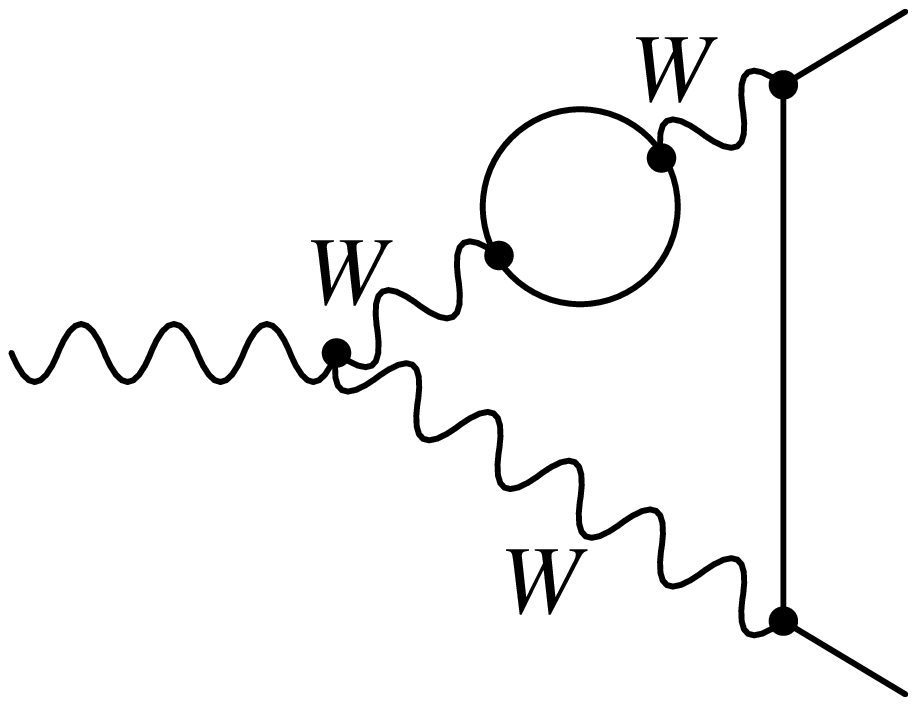, width=3.9cm}\\[1ex]
\raisebox{3cm}{(c)}\psfig{figure=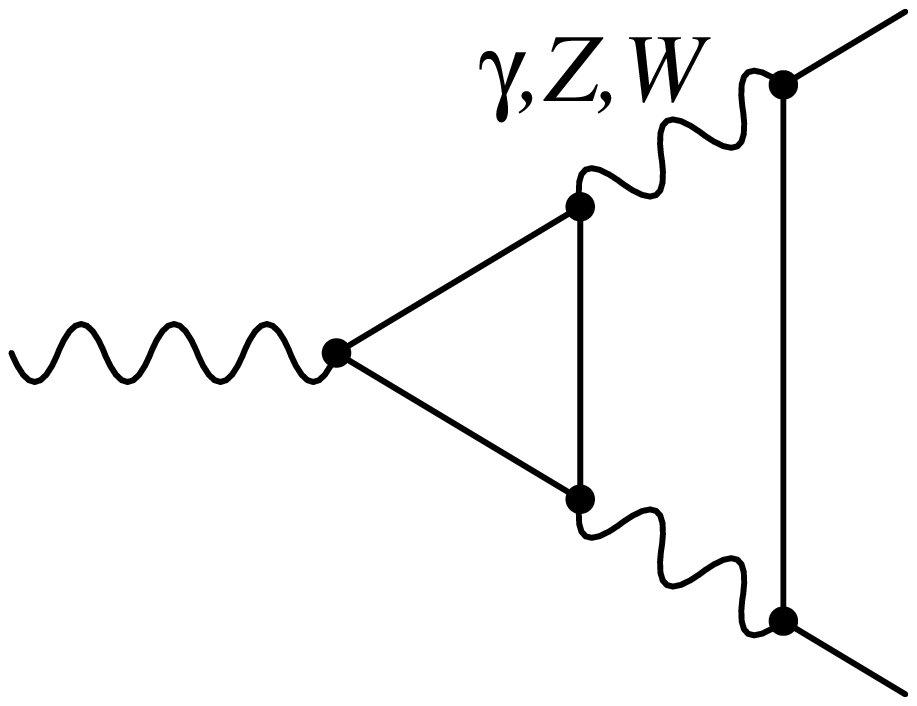, width=3.9cm} \hspace{10mm}
\raisebox{3cm}{(d)}\psfig{figure=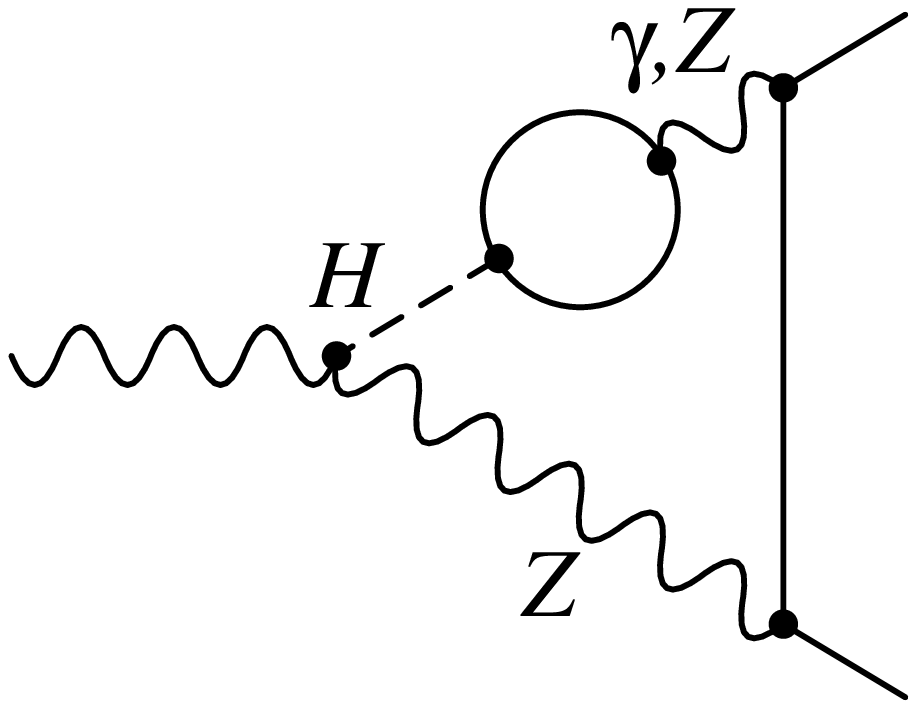, width=3.9cm}
\end{center}
\caption{
Genuine fermionic two-loop $Zl^+l^-$ vertex diagrams contributing to \SinEff.
\label{fig:diags}}
\end{figure}

\begin{figure}[tb]
\begin{center}
\raisebox{3cm}{(a)}\psfig{figure=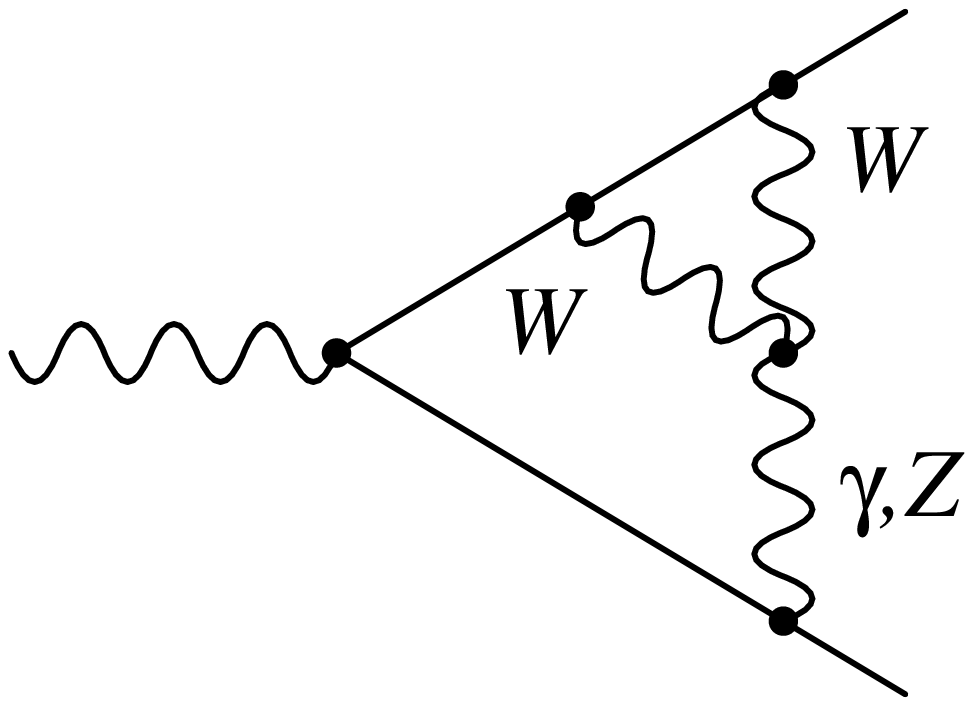, width=4.5cm} \hspace{6mm}
\raisebox{3cm}{(b)}\psfig{figure=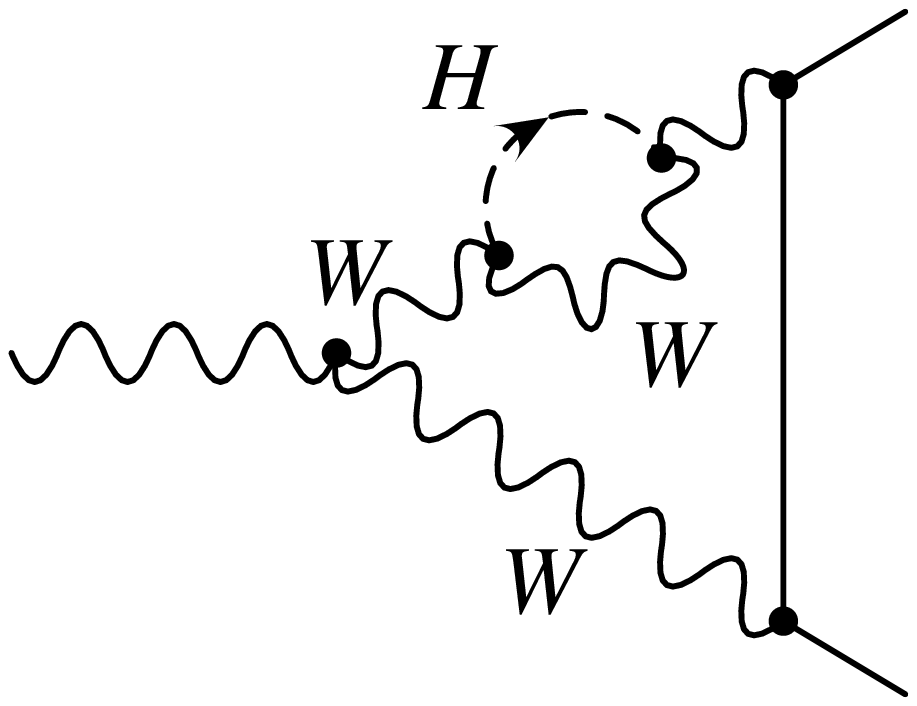, width=3.9cm}\\[1ex]
\rule{0mm}{0mm}\hspace{3mm}
\raisebox{3cm}{(c)}\psfig{figure=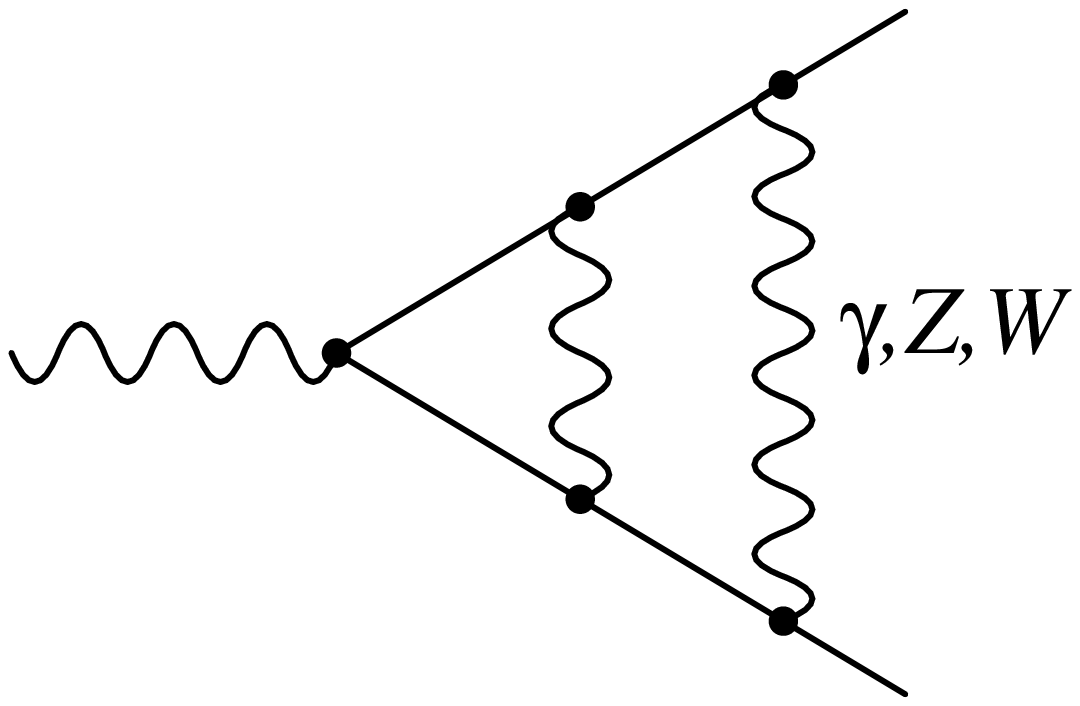, width=5cm} \hspace{1mm}
\raisebox{3cm}{(d)}\psfig{figure=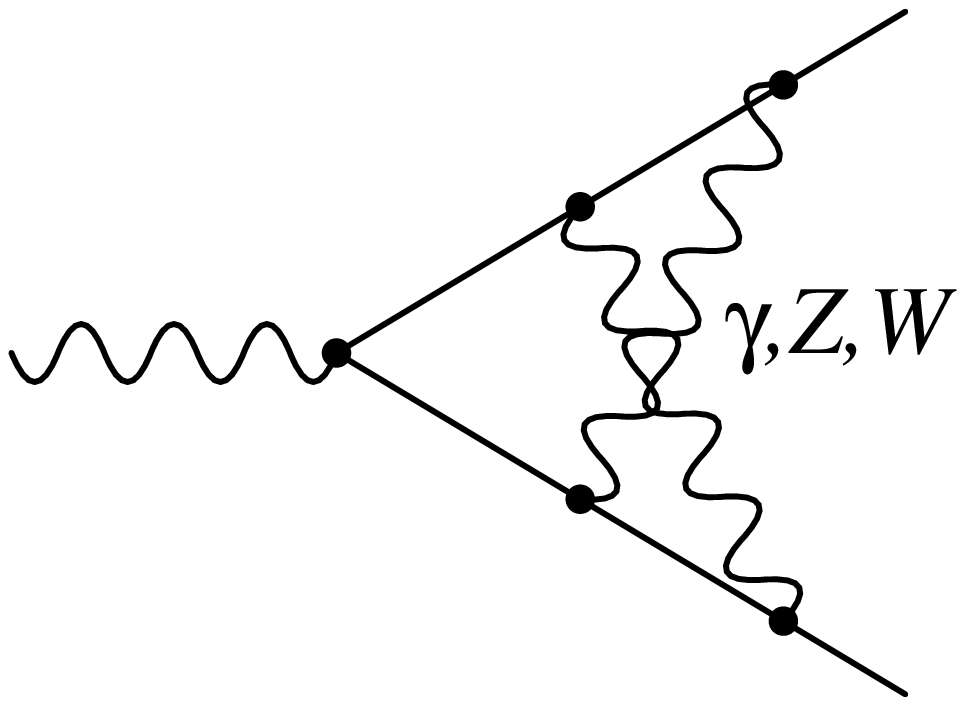, width=4.3cm}
\end{center}
\caption{
Examples of bosonic two-loop $Zl^+l^-$ vertex diagrams contributing to \SinEff.
\label{fig:diagsb}}
\end{figure}

This paper discusses the complete computation of all electroweak two-loop
corrections to \SinEff.
In addition to the corrections to the prediction of the
$W$-boson mass, which have been analyzed before \cite{muon,muon2}, this
includes all two-loop diagrams contributing to the $Zl^+l^-$ vertex on the $Z$
pole. The diagrams can be conveniently divided into two groups; {\it fermionic}
contributions with at least one closed fermion loop, and {\it bosonic}
contributions without closed fermion loops.
The genuine fermionic two-loop vertex 
diagrams are represented by the generic topologies in Fig.~\ref{fig:diags} and
some examples of bosonic two-loop diagrams are given in Fig.~\ref{fig:diagsb}.

Results for the complete two-loop corrections have been presented first in Ref.~\cite{sineff,sinbos}. The results
for the fermionic contributions  have been confirmed in Ref.~\cite{fermionic2}
and partial results for the bosonic contributions were also obtained in
Ref.~\cite{Hollik:2005ns}. This paper describes the computational methods and
analysis in more detail.

The paper is organized as follows. In section~\ref{sc:calc}, the process
$e^+e^- \to l^+l^-$ is analyzed at next-to-next-to-leading order near the
$Z$-boson pole and the $\Oaa$ definition of the \SinEff\ is extracted.
Furthermore the general strategies for the calculation of two-loop
contributions to the form factor $\Delta\kappa$ are discussed.
Sections~\ref{sc:2loop} and \ref{sc:2bos} explain the calculation of
the fermionic and bosonic two-loop diagrams in
detail. For two-loop vacuum and self-energy diagrams, well-established
techniques exist and have been used for the computation of $\MW$
\cite{muon,muon2,AC2}. The new part in this project are the two-loop vertex
topologies, which have been treated with two conceptually independent methods.
A discussion of the numerical results and remaining theoretical uncertainties
due to unknown higher orders can be found in section~\ref{sc:results}. In
addition to the effective leptonic weak mixing angle, results are given also
for the effective weak mixing angle for other final state flavors, i.e.\ for
couplings of the $Z$ boson to other fermions. Finally the implementation of our
new results into the program {\sc Zfitter} is described.


\section{Outline of the calculation}
\label{sc:calc}

The two-loop corrections to the effective weak mixing angle \SinEfff\ are part
of the next-to-next-to-leading order corrections to the process $e^+e^- \to f
\bar{f}$ for center-of-mass energies near the $Z$-boson mass, $\sqrt{s} \approx
\MZ$. To set the scene for this calculation,  a framework for the
next-to-next-to-leading order analysis of $f \bar{f}$ production needs to be
established. Furthermore it has to be checked whether \SinEfff\ is a
well-defined, i.e.\ gauge-invariant and finite, quantity at this order in
perturbation theory.

\subsection{Definition of the effective weak mixing angle at
next-to-next-to-leading order}

In higher-order calculations, occurrences of unstable intermediate particles
need to be treated carefully in order to preserve gauge-invariance and
unitarity. Currently, the only scheme proven to fulfill both requirements 
to all orders in perturbation theory is the
{\it pole scheme} \cite{unstab,hvelt,unstab2}. 
It involves a systematic Laurent expansion
around the complex pole ${\cal M}^2 = M^2 - i M \Gamma$ associated with the
propagator of the unstable particle with mass $M$ and width $\Gamma$. In the
case of the process $e^+e^- \to f \bar{f}$, $e \neq f$, near the $Z$ pole, the amplitude is
written as
\begin{equation}
{\cal A}[e^+e^- \to f \bar{f}] = \frac{R}{s-{\cal M}_\PZ^2} + S + 
	(s-{\cal M}_\PZ^2) S' +
\dots \label{eq:polexp}
\end{equation}
with
\begin{equation}
{\cal M}_\PZ^2 = \mz^2 -  i \mz \gz.
\end{equation}
Owing to the analyticity of the S-matrix, all coefficients of Laurent expansion,
$R,S,S',\dots$ and the pole location ${\cal M}_\PZ^2$
are individually gauge-invariant, UV- and IR-finite, when soft and collinear real photon emission is added.

The first term in \eqref{eq:polexp} corresponds to a Breit-Wigner
parametrization of the $Z$ line shape with a constant decay width.
Experimentally, however, the gauge-boson mass is determined based on a
Breit-Wigner function with a running (energy-dependent) width,
\begin{equation}
{\cal A} \propto \frac{1}{s - \MZ^2 + i s \GZ / \MZ}.
\end{equation}
As a consequence of these different parameterizations, there is a shift between
the experimental mass parameter, $\MZ$, and the mass parameter of the pole
scheme, $\mz$, \cite{massshift},
\begin{equation}
\mz^2 = \MZ^2 / (1+ \GZ^2/\MZ^2),
\end{equation}
amounting to $\mz \approx \MZ - 34.1$ MeV. In the following, barred quantities
always refer to pole scheme parameters.

The evaluation of higher order contributions in the pole scheme
involves a simultaneous expansion around the pole location and in the
perturbation order $\alpha$. Since near the $Z$ pole $\alpha$, $\GZ$ and
$(s-{\cal M}_\PZ^2)$ are all of the same order, for a next-to-next-leading order
calculation $R$ needs to be determined to \Oaa, $S$ only to ${\cal
O}(\al)$, while a tree-level result is sufficient for $S'$.

The effective weak mixing angle is contained in the pole term residue $R$ in
\eqref{eq:polexp}. For further use, the following notations for vertex and
self-energy form factors are introduced,
\begin{align}
\raisebox{-9mm}{\psfig{figure=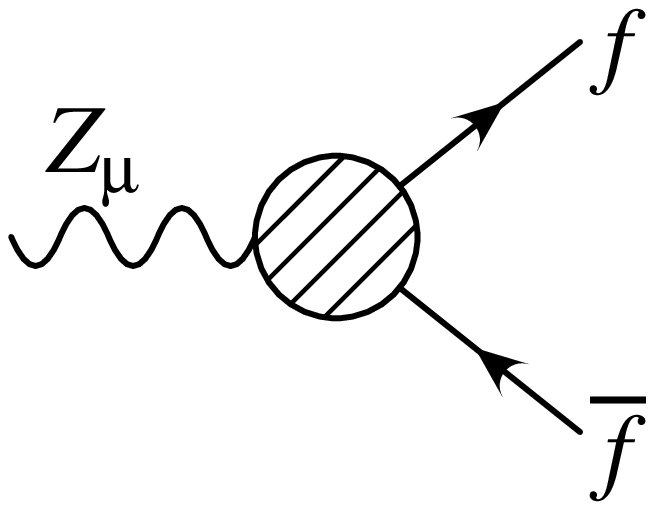, width=2.5cm}}
 &\equiv \Gamma[Z_\mu f\bar{f}] \equiv
 z_{\Pf,\mu} = i \gamma_\mu (v_\Pf+a_\Pf \gamma_5), \\
\raisebox{-9mm}{\psfig{figure=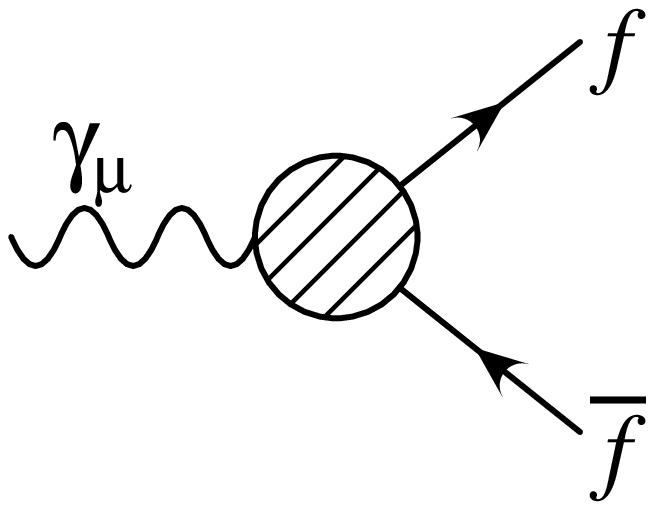, width=2.5cm}}
 &\equiv \Gamma[\gamma_\mu f\bar{f}] \equiv
 g_{\Pf,\mu} = i \gamma_\mu (q_\Pf+p_\Pf \gamma_5), \\
\raisebox{-2mm}{\psfig{figure=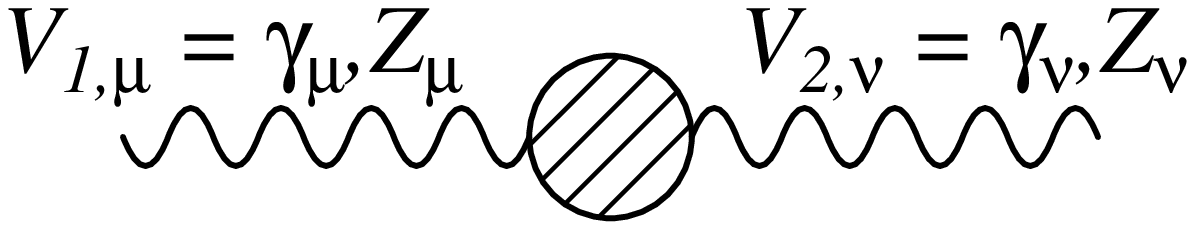, width=4cm}}
 &= \Sigma_{V_1 V_2}^{\mu\nu},
\end{align}
where the shaded blobs stand for one-particle irreducible loop contributions.
It is also convenient to define $Zf\bar{f}$ vertex form factors including the
effect of $Z$-$\gamma$ mixing,
\begin{equation}
\begin{aligned}
\hat{z}_{\Pf,\mu}(k^2) &= i \gamma_\mu 
\left[\hat{v}_\Pf(k^2)+\hat{a}_\Pf(k^2) \gamma_5\right] \\
&\equiv i \gamma_\mu \left[v_\Pf(k^2)+a_\Pf(k^2) \gamma_5\right] 
     - i \gamma_\mu \left[q_\Pf(k^2)+p_\Pf(k^2) \gamma_5\right]
\frac{\Sigma_{\gamma Z}(k^2)}{k^2 + \Sigma_{\gamma \gamma}(k^2)} \\
&=
\raisebox{-9mm}{\psfig{figure=Zvff.ps, width=2.5cm}} +
\raisebox{-9mm}{\psfig{figure=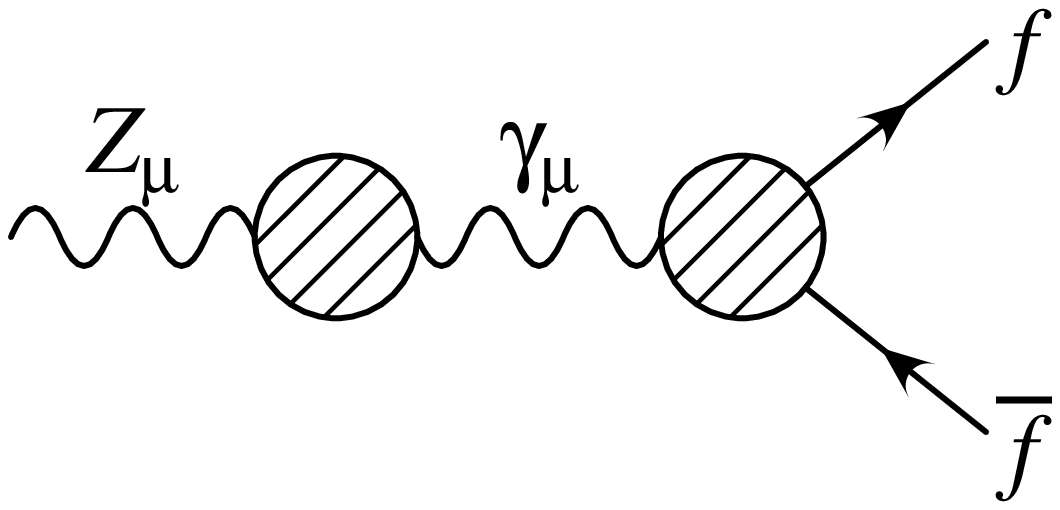, width=3.8cm}} +
\raisebox{-9mm}{\psfig{figure=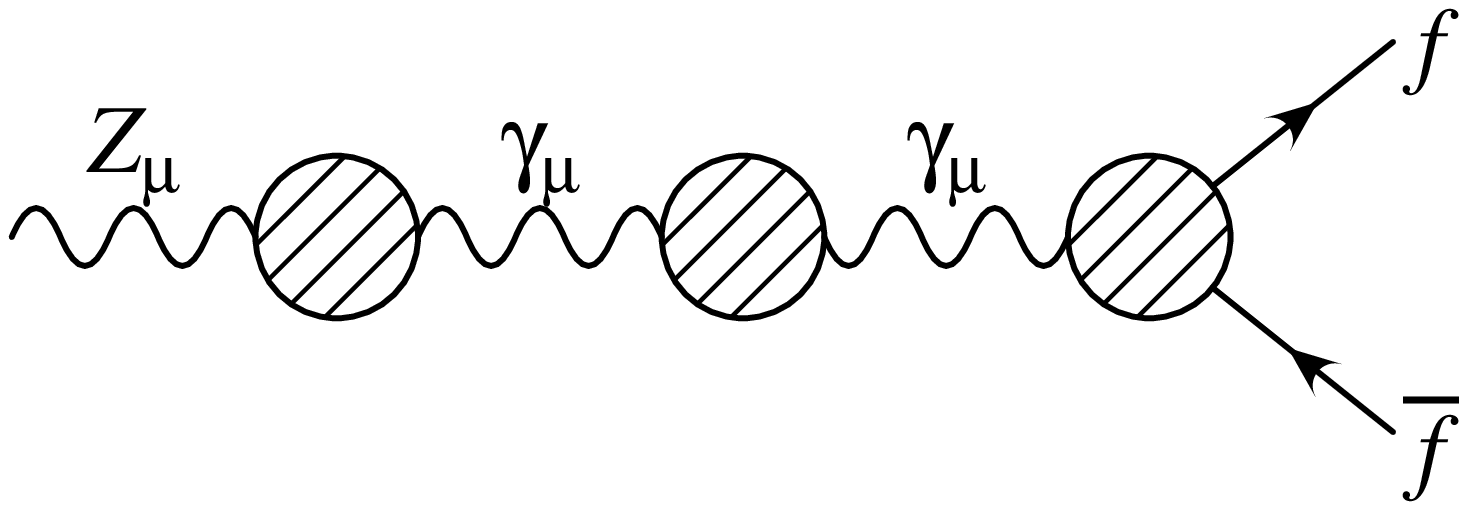, width=5.1cm}} + \dots,
\end{aligned}
\end{equation}
where $k$ is the momentum of the external $Z$ line.
With these definitions, the residue $R$ up to next-to-next-to-leading order can
be cast into the form \cite{hvelt}
\begin{align}
R &= 
 \begin{aligned}[t]
 &z_\Pe^{(0)} \, R_{ZZ} \, z_\Pf^{(0)} + \left[ \hat{z}_\Pe^{(1)}(\MZ^2)\, z_\Pf^{(0)} +
   z_\Pe^{(0)} \, \hat{z}_\Pf^{(1)}(\MZ^2) \right] 
   \left[ 1+ {\Sigma_{\gamma Z}^{(1)}}'(\MZ^2) \right] \\
 &+ \hat{z}_\Pe^{(2)}(\MZ^2) \, z_\Pf^{(0)} +
   z_\Pe^{(0)} \, \hat{z}_\Pf^{(2)}(\MZ^2)
   + \hat{z}_\Pe^{(1)}(\MZ^2) \, \hat{z}_\Pf^{(1)}(\MZ^2) \\
 &- i \MZ \GZ \left[ \mbox{$\hat{z}_\Pe^{(1)}$}'(\MZ^2)\, z_\Pf^{(0)} +
   z_\Pe^{(0)} \, \mbox{$\hat{z}_\Pf^{(1)}$}'(\MZ^2) \right],
 \end{aligned} \\
R_{ZZ} &= 
 \begin{aligned}[t]
 &1 - {\Sigma_{Z Z}^{(1)}}'(\MZ^2) \\
 &- {\Sigma_{Z Z}^{(2)}}'(\MZ^2) + \left( {\Sigma_{Z Z}^{(1)}}'(\MZ^2) \right)^2
   + i \MZ \GZ \, {\Sigma_{Z Z}^{(1)}}''(\MZ^2) \\
 &- \frac{1}{\MZ^4} \left( {\Sigma_{\gamma Z}^{(1)}}(\MZ^2) \right)^2
   + \frac{2}{\MZ^2} \, {\Sigma_{\gamma Z}^{(1)}}(\MZ^2) \,
     {\Sigma_{\gamma Z}^{(1)}}'(\MZ^2).
 \end{aligned}
\end{align}
Here the Lorentz indices have been suppressed. Based on the definition of
\SinEff\ in eqs.~\eqref{eq:def},\eqref{eq:kappa}, the two-loop result of the
effective weak mixing angle is derived from $R$ as
\begin{equation}
\begin{aligned}
\sinefff &\equiv \left(1- \frac{\mw^2}{\mz^2} \right) 
	\re\! \left\{ 1 + \De\overline{\kappa}_\PZ^\Pf(\MZ^2) \right\} \\
 &= \left(1- \frac{\mw^2}{\mz^2} \right) \re\Biggl\{ 1+
   \frac{\hat{a}_\Pf^{(1)} \, v_\Pf^{(0)} - 
   	\hat{v}_\Pf^{(1)} \, a_\Pf^{(0)}}{a_\Pf^{(0)}(a_\Pf^{(0)}-v_\Pf^{(0)})}
	\Biggr|_{k^2 = \MZ^2} \\
 &\hspace{6em}
     + \frac{\hat{a}_\Pf^{(2)} \, v_\Pf^{(0)} \, a_\Pf^{(0)} - 
   	\hat{v}_\Pf^{(2)} \, (a_\Pf^{(0)})^2 -
	(\hat{a}_\Pf^{(1)})^2 \, v_\Pf^{(0)} +
	\hat{a}_\Pf^{(1)} \, \hat{v}_\Pf^{(1)} \, a_\Pf^{(0)}}{(a_\Pf^{(0)})^2
	(a_\Pf^{(0)}-v_\Pf^{(0)})} \Biggr|_{k^2 = \MZ^2}
   \Biggr\}.
\end{aligned} \label{eq:sin2}
\end{equation}
Since the pole scheme is based on a formal Laurent series of the physical
amplitude, all coefficients in the expansion and thus the effective weak mixing
angle are manifestly gauge-invariant and UV-finite. While the pole scheme
formalism does not make any statement about IR finiteness, it can be checked
that eq.~\eqref{eq:sin2} is also a IR-safe quantity, 
i.e.\ all IR-divergencies from photon exchange
diagrams cancel. Similarly, collinear divergencies (or Sudakov factors for massive 
fermions) also cancel. This can be explained by the fact that the QED
contributions in the soft and collinear limits factorize from massive loop effects and
therefore drop out in the ratio of the vector and axial-vector form factor in
eq.~\eqref{eq:def}. At the diagrammatic level, this cancellation of
divergencies occurs not only between two-loop diagrams, but also between
2-loop and products of 1-loop diagrams, for example
\begin{equation}
\raisebox{-10mm}{\psfig{figure=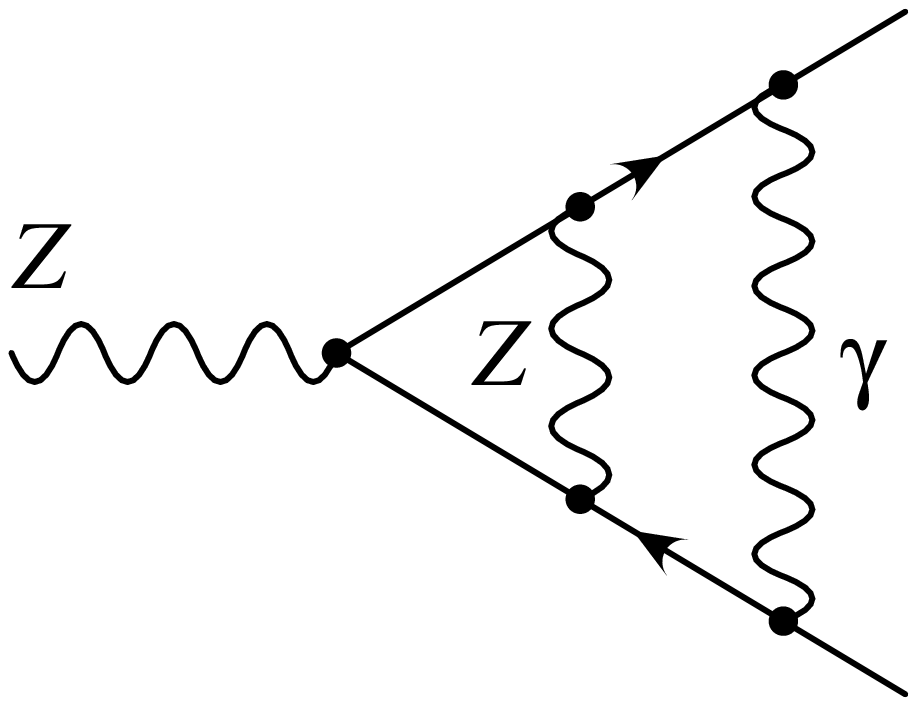, width=3cm}}
 \;\;=\;\; \raisebox{-7mm}{\psfig{figure=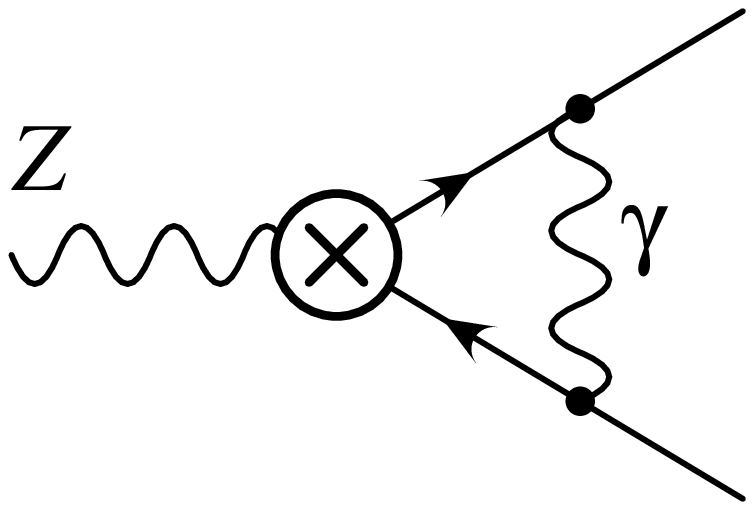, width=2.4cm}} + \mbox{finite},
 \qquad \mbox{with} \qquad \otimes = \raisebox{-7mm}{\psfig{figure=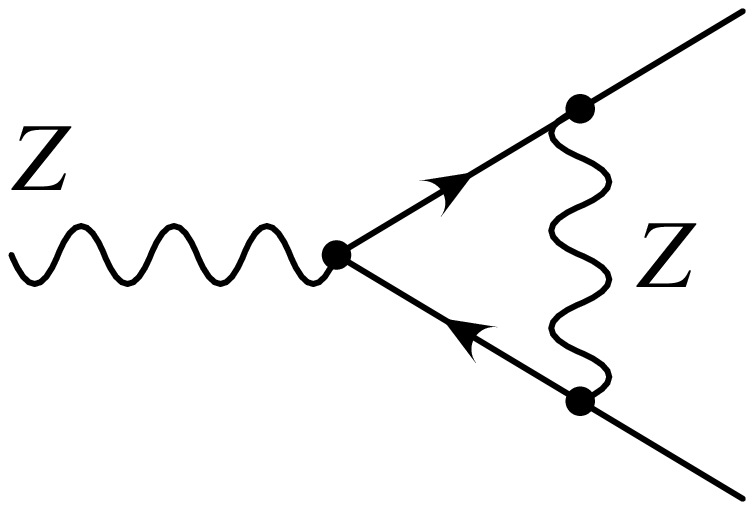, width=2.4cm}}
 \;.
\end{equation}

Experimentally, the effective weak mixing angle is determined from measurements
of forward-backward and left-right asymmetries of the process $e^+e^- \to f
\bar{f}$. The derivation of \SinEfff\ from these asymmetries requires the
subtraction of effects from QED and QCD corrections, s-channel photon
exchange and $\gamma$-$Z$ interference, off-shellness of the $Z$-boson and box
contributions. These non-resonant effects enter into the amplitude through the
next-to-leading term $S$ in the pole expansion \eqref{eq:polexp}, and need to be
computed up to one-loop order. In order to
relate the $\Oaa$ result \eqref{eq:sin2} for \SinEfff\ to the
value quoted by the experimental analyses, it needs to be checked that the
subtracted effects are consistent with the pole scheme prescription.

In experimental studies, the program {\sc Zfitter} \cite{zfitter} is widely used
for prediction of the contributions from QED and QCD corrections, s-channel photon
exchange and $\gamma$-$Z$ interference, off-shellness of the $Z$-boson and box
contributions. In {\sc Zfitter}, the radiative corrections to the process $e^+e^- \to f
\bar{f}$ are parametrized by four form factors $\rho_{\Pe\Pf}$, $\kappa_{\Pe}$,
$\kappa_{\Pf}$, $\kappa_{\Pe\Pf}$,
\begin{equation}
\begin{aligned}
{\cal A}[e^+e^- \to f \bar{f}] = \;&4\pi i\,\alpha \, \frac{Q_\Pe Q_\Pf}{s} \,
 \gamma_\mu \otimes \gamma^\mu \\
&+ i \frac{\sqrt{2} \GF \MZ^2}{1 + i \GZ/\MZ} \, I_\Pe^{(3)} \, I_\Pf^{(3)} \,
  \frac{1}{s - \mz^2 + i \mz \gz} \\
& \quad \times \rho_{\Pe\Pf} \, \Bigl[
	\gamma_\mu(1+\gamma_5) \otimes \gamma^\mu(1+\gamma_5) \\
& \qquad\qquad - 4 |Q_\Pe| \sw^2 \, \kappa_{\Pe} \, \gamma_\mu \otimes
  \gamma^\mu(1+\gamma_5) \\[1ex]
& \qquad\qquad - 4 |Q_\Pf| \sw^2 \, \kappa_{\Pf} \, \gamma_\mu(1+\gamma_5) \otimes
  \gamma^\mu \\
& \qquad\qquad + 16 |Q_\Pe Q_\Pf| \sw^4 \, \kappa_{\Pe\Pf} \, \gamma_\mu \otimes
  \gamma^\mu \Bigr]
\end{aligned}
\end{equation}
Note that apart from the $Z$ propagator, the gauge boson masses are defined
according to the running width prescription (un-barred symbols) instead of the
pole scheme definition (barred symbols).
As a result the form factors $\kappa_{\Pe}$, $\kappa_{\Pf}$, $\kappa_{\Pe\Pf}$
can differ from the corresponding form factors $\overline{\kappa_{\Pe}}$,
$\overline{\kappa_{\Pf}}$, $\overline{\kappa_{\Pe\Pf}}$ in the pole scheme. In
the following, the relation between the two sets of quantities will be worked
out.

{\sc Zfitter} includes all radiative corrections to $e^+e^- \to f \bar{f}$
consistently at the one-loop level with some leading two-loop contributions.
However, it has not been designed for a complete next-to-next-to-leading order
analysis and inconsistencies could occur at this level. In {\sc Zfitter} QED
and QCD corrections are included via a convolution of the cross-section. They
will be discussed in more detail later. The effects from s-channel photon
exchange, $\gamma$-$Z$ interference, off-shellness of the $Z$-boson and massive
(non-QED) box contributions are taken into account by the
formulae~\cite{zfitter}
\begin{align}
\kappa_{\Pe\Pf}(s) &= \kappa_{\Pe}(s) \kappa_{\Pf}(s) - \frac{\MZ^2-s}{s} \;
  \frac{1}{(a^{(0)}_\Pe - v^{(0)}_\Pe)(a^{(0)}_\Pf - v^{(0)}_\Pf)} 
  \nonumber \\
& \qquad \times \left[ q^{(1)}_\Pe q^{(0)}_\Pf +
			 q^{(1)}_\Pf q^{(0)}_\Pe -
			 p^{(1)}_\Pf q^{(0)}_\Pe \frac{v^{(0)}_\Pf}{a^{(0)}_\Pf}
		       - p^{(1)}_\Pe q^{(0)}_\Pf \frac{v^{(0)}_\Pe}{a^{(0)}_\Pe}
		       - q^{(0)}_\Pe q^{(0)}_\Pf 
		       \frac{\Sigma^{(1)}_{\gamma\gamma}}{s} +  
		       \mbox{boxes} \right], \label{eq:kap1} \\
\kappa_{\Pe,\Pf}(s) &= \kappa_\PZ^{\Pe,\Pf}(s) + \frac{\MZ^2-s}{s}
 \left[ \frac{q^{(0)}_{\Pe,\Pf}}{a^{(0)}_{\Pe,\Pf} - v^{(0)}_{\Pe,\Pf}} \,
 	\frac{p^{(1)}_{\Pf,\Pe}}{a^{(0)}_{\Pf,\Pe}} + \mbox{boxes} \right], \label{eq:kap2} \\
\kappa_\PZ^\Pf(s) &= \kappa_\PZ^\Pf(\MZ^2) + (s -\MZ^2)\,
  \frac{\mbox{$\hat{a}_\Pf^{(1)}$}'(\MZ^2) \, v_\Pf^{(0)} - 
   	\mbox{$\hat{v}_\Pf^{(1)}$}'(\MZ^2) \,
	a_\Pf^{(0)}}{a_\Pf^{(0)}(a_\Pf^{(0)}-v_\Pf^{(0)})}.
\end{align}
From the pole expansion scheme one obtains in contrast to
eqs.~\eqref{eq:kap1},\eqref{eq:kap2}
\begin{align}
\overline{\kappa}_{\Pe\Pf}(s) &= \overline{\kappa}_{\Pe}(s) \overline{\kappa}_{\Pf}(s) 
	- \frac{\MZ^2-i\MZ\GZ-s}{s} \;
  \frac{1}{(a^{(0)}_\Pe - v^{(0)}_\Pe)(a^{(0)}_\Pf - v^{(0)}_\Pf)} 
  \nonumber \\
& \qquad \times \left[ q^{(1)}_\Pe q^{(0)}_\Pf +
			 q^{(1)}_\Pf q^{(0)}_\Pe -
			 p^{(1)}_\Pf q^{(0)}_\Pe \frac{v^{(0)}_\Pf}{a^{(0)}_\Pf}
		       - p^{(1)}_\Pe q^{(0)}_\Pf \frac{v^{(0)}_\Pe}{a^{(0)}_\Pe}
		       - q^{(0)}_\Pe q^{(0)}_\Pf 
		       \frac{\Sigma^{(1)}_{\gamma\gamma}}{s} +  
		       \mbox{boxes} \right],  \\
\overline{\kappa}_{\Pe,\Pf}(s) &= \overline{\kappa}_\PZ^{\Pe,\Pf}(s) + \frac{\MZ^2-i\MZ\GZ-s}{s}
 \left[ \frac{q^{(0)}_{\Pe,\Pf}}{a^{(0)}_{\Pe,\Pf} - v^{(0)}_{\Pe,\Pf}} \,
 	\frac{p^{(1)}_{\Pf,\Pe}}{a^{(0)}_{\Pf,\Pe}} + \mbox{boxes} \right]. 
	\label{eq:kapb2}
\end{align}
with
\begin{align}
\overline{\kappa}_{\Pf} &= \kappa_{\Pf} \left[1+ \frac{\cw^2}{\sw^2}
  \left(\frac{\GW^2}{\MW^2} - \frac{\GZ^2}{\MZ^2} \right)\right], \\
\overline{\kappa}_{\Pe\Pf} &= \kappa_{\Pe\Pf} \left[1+ \frac{\cw^2}{\sw^2}
  \left(\frac{\GW^2}{\MW^2} - \frac{\GZ^2}{\MZ^2} \right)\right]^2,
\end{align}
Note that for next-to-next-to-leading accuracy it is not necessary to
distinguish between barred and un-barred symbols in the radiative corrections,
since $\mz^2 - \MZ^2 = \Oaa$.

From eqs.~(\ref{eq:kap1}--\ref{eq:kapb2}) one finds a difference for the
derivation of the value of \SinEfff\ between {\sc Zfitter} and the pole scheme:
\begin{align}
{\sinefff}_{\mbox{\scriptsize\sc ,Zfitter}} &= \sw^2 \, 
	\re\! \left\{ \kappa_\PZ^\Pf(\MZ^2) \right\} \\
{\sinefff}_{\rm ,pole} &= \overline{s}_{\scrs\PW}^2 \,
	\re\! \left\{ \overline{\kappa}_\PZ^\Pf(\MZ^2) \right\}
  = {\sinefff}_{\mbox{\scriptsize\sc ,Zfitter}} 
    - \frac{\GZ}{\MZ} \, \frac{q^{(0)}_\Pf}{a^{(0)}_\Pe
    	(a^{(0)}_\Pf - v^{(0)}_\Pf)} \; \im \!\! \left\{ p^{(1)}_\Pe \right\} 
  \label{eq:shift}
\end{align}
with 
\begin{equation}
\overline{s}_{\scrs\PW}^2 = \left(1- \frac{\mw^2}{\mz^2} \right) =
  \sw^2 \left[1+ \frac{\cw^2}{\sw^2}
  \left(\frac{\GW^2}{\MW^2} - \frac{\GZ^2}{\MZ^2} \right)\right]^{-1}.
\end{equation}
A similar deviation is found for the contribution of the form factors
$\kappa_{\Pe\Pf},\overline{\kappa}_{\Pe\Pf}$ between the two schemes, which
however cannot be expressed directly as a shift in \SinEfff.

In principle, an additional discrepancy arises from the box contributions.
The massive boxes with $Z$ and $W$ boson exchange are included in {\sc Zfitter}
at the one-loop level, which is sufficient for the next-to-next-to-leading order
calculation in the pole scheme. Nevertheless, in \eqref{eq:kapb2} an extra
term stemming from the box contributions arises, which is proportional to
$i\MZ\GZ$. However, this term does not 
contribute to the squared matrix element since the massive boxes have no
absorptive part\footnote{A special case is Bhabha scattering, $f=e$, where
additional box and
    t-channel diagrams contribute. For the purpose of this work, the subtraction
    of these contributions has not been analyzed in detail, justified by the
    fact that the $e^+e^-$ final state has a relatively small impact on the
    determination of the effective weak mixing angle at present. In general, a
    more careful analysis of this process should be done in the future.}.

Besides the contributions from s-channel photon exchange and boxes, the
translation between the measured asymmetries and the effective weak mixing angle
requires the subtraction of QED and QCD corrections to the external fermions. 

In the left-right asymmetry, the effect of final state QED and QCD corrections
and initial-final state QED interference cancels \cite{QEDLR} up to
next-to-next-to-leading order. Initial-state QED radiation can be treated
through convolution with a radiator function and has been computed including
the exact $\Oaa$ corrections and higher-order leading contributions
\cite{ISRasq}.

For the forward-backward asymmetry on the $Z$ pole, the contribution from
final-state virtual and soft photon radiation vanishes for massless external
fermions \cite{QEDLR, QEDFB, hollikee}. This statement is valid up to
corrections of the order ${\cal O}(\al \, \Delta E_\gamma/\sqrt{s})$, where
$\Delta E_\gamma$ is the soft-photon cut-off, and terms of order ${\cal
O}(\al\, \mf/\sqrt{s}$, where $\mf$ is the final-state fermion mass.
Nevertheless, the complete one-loop contributions to final-state radiation are
known and taken into account in the extraction of the effective weak mixing
angle \cite{zfitter}. The leading effect of final-state fermion masses of
${\cal O}(\al \, \mf/\sqrt{s})$ is also known and included \cite{QEDFBfinmass},
with the remaining effects of order ${\cal O}(\al^2 \Delta E_\gamma/\sqrt{s})$,
${\cal O}(\al^2 \mf/\sqrt{s}$, ${\cal O}(\al \, \mf^2/s)$ being numerically
negligible for the two-loop analysis for \SinEfff\  under study here. Multiple
hard final-state photon radiation is taken into account by Monte-Carlo methods,
see e.g. \cite{koralz}, with a small numerical error. QCD final state effects
are treated similarly to the QED contributions.

Interference of initial-final state photon radiation is also known up to order
\Oa for the forward-backward asymmetry. For sufficiently loose soft-photon cut,
$\Delta E_\gamma \gsim \GZ$, the initial-final interference of soft and virtual
photons at the $Z$ pole is suppressed by the width $\GZ$ of the $Z$ boson
\cite{QEDFB,hollikee}, so that the $\Oaa$ contribution is effectively of order ${\cal
O}(\al^2 \GZ/\MZ)$, i.e.\ beyond the next-to-next-to-leading order corrections
under study in this work. As before, initial-state radiation to the
forward-backward asymmetry is included up to \Oaa, and partially beyond, by
means of a convolution. Thus while a complete next-to-next-to-leading order
calculation of QED corrections to the forward-backward asymmetry is not
available, the present treatment of QED corrections is sufficient for a two-loop
analysis of \SinEfff.
Nevertheless, a complete $\Oaa$ calculation of QED effects would be desirable.

In summary, it was found that the treatment of non-resonant contributions in
{\sc Zfitter} is not consistent with the pole scheme at next-to-next-to-leading
order. As a result, the value of \SinEfff\ needs to be corrected by a shift
\begin{equation}
\sw^2 \, \delta\kappa_\Pf = - \frac{\GZ}{\MZ} \, \frac{q^{(0)}_\Pf}{a^{(0)}_\Pe
    	(a^{(0)}_\Pf - v^{(0)}_\Pf)} \; \im \!\! \left\{ p^{(1)}_\Pe \right\}.
\end{equation}
Numerically this shift amounts to $\sw^2 \, \delta\kappa_\Pf \approx 1.5 \times
10^{-6}$, well below the current experimental error of $1.7 \times 10^{-4}$
\cite{exp}.
Therefore, this shift will be neglected in the analysis in
section~\ref{sc:results}. It was checked that a similar shift $\delta
\kappa_{\Pe\Pf}$ in the form factor $\kappa_{\Pe\Pf}$ also leads to a negligible
numerical effect on \SinEfff.

\subsection{Renormalization}

In this work the on-shell renormalization scheme is employed. It defines the
mass parameters and coupling constants in close relation to physical
observables. The renormalized squared masses are defined as the real part of
the propagator poles, while the external fields are
renormalized to unity at the position of the poles. The electromagnetic charge
is defined as the coupling strength of the electromagnetic vertex in the Thomson
limit.
Explicit expressions for the necessary counterterms can be found in
Ref.~\cite{muon2}.

As described in the previous section, the computation of radiative corrections
to the effective weak mixing angle entails the calculation of loop
contributions to the $Z f\bar{f}$ vertex. In principle this involves a field
renormalization for the $Z$ boson, which appears as an external particle of
the vertex. Beyond one-loop order, the treatment of field renormalizations for
unstable particles proves to be not straightforward~\cite{nkrren}. However, in
the calculation of \SinEfff\ all occurrences of the $Z$ boson field
renormalization drop out between the vector and axial-vector form factors in
eq.~\eqref{eq:sin2}. The independence of \SinEfff\ on the total normalization
of the $Z$ boson field strength can already be seen in  eq.~\eqref{eq:def},
where the effective weak mixing angle is defined through the ratio of vertex
form factors.

While the on-shell counterterms cancel the UV-divergencies in the virtual loop
corrections, all IR- and collinear divergencies drop out in the quantity
\SinEfff, as explained in the previous section. The computation of the loop
integrals is performed using dimensional regularization. With this
regularization scheme, special care is needed for the treatment of the
$\gamma_5$ matrix in triangle fermion sub-loops. A practical solution to this
problem will be discussed in detail in section \ref{sc:gamma5}.

\subsection{Preliminaries}

Throughout the calculation of the two-loop corrections, the masses and Yukawa
couplings of all fermions but the top quark are neglected. The quark mixing
matrix is assumed to be diagonal. The vector and axial-vector components of the
vertex corrections $\hat{z}_{\Pf,\mu}$ were projected out by contraction with 
suitable projection operators,
\begin{align}
\hat{v}_\Pf(k^2) &= \frac{1}{2(2-D)k^2} \, {\rm Tr}[\gamma^\mu \, \pslash_1 \,
\hat{z}_{\Pf,\mu}(k^2) \, \pslash_2], \\
\hat{a}_\Pf(k^2) &= \frac{1}{2(2-D)k^2} \, {\rm Tr}[\gamma_5 \,
\gamma^\mu \, \pslash_1 \,
\hat{z}_{\Pf,\mu}(k^2) \, \pslash_2], 
\end{align}
where $D$ is the space-time dimension and $p_{1,2}$ are the momenta of the
external fermions. As a result, only scalar integrals remain after projection,
but there are non-trivial structures of scalar products in the numerators 
of the integrals, which require further treatment.


\section{Calculation of fermionic two-loop vertex diagrams}
\label{sc:2loop}

The computation of the two-loop corrections to the effective weak mixing angle
can be divided into the calculation of the vertex loop contributions to the $Z
f\bar{f}$ vertex and the on-shell counterterms. The latter involve two-loop
vacuum and self-energy contributions, similar to the two-loop corrections to
the $W$-boson mass~\cite{muon,muon2}, while the former also contain two-loop
vertex topologies as a new complication. The generic two-loop vertex diagrams
with closed fermion loops are shown in Fig.~\ref{fig:diags}.

The evaluation of the two-loop vertex contributions has been performed with
two independent methods, in order to allow for a
non-trivial check of the result. One method is based on large mass expansions
for the diagrams involving internal top quark propagators and differential
equations for the diagrams with only light fermions. The second method makes use
of numerical integrations derived from dispersion relations and Feynman
parameterizations.

\subsection{Large top-quark mass expansions and analytical results}
\label{sc:largemt}

This approach divides the fermionic two-loop vertices in two categories:
diagrams with internal top quark lines and diagrams that have only light fermion
lines.

Observing that the ratio $x = \MZ^2/\mt^2 \sim 1/4$ is a small number, the
top-quark contributions can be conveniently calculated by performing an
expansion in $x$. The coefficients of this large-mass expansion decompose
completely into one-loop integrals and two-loop vacuum integrals, for which
analytical formulae are available in the literature \cite{davtausk}. 

\begin{figure}[tb]
\rule{0mm}{0mm}\hfill
\begin{minipage}[b]{5cm}
\psfig{figure=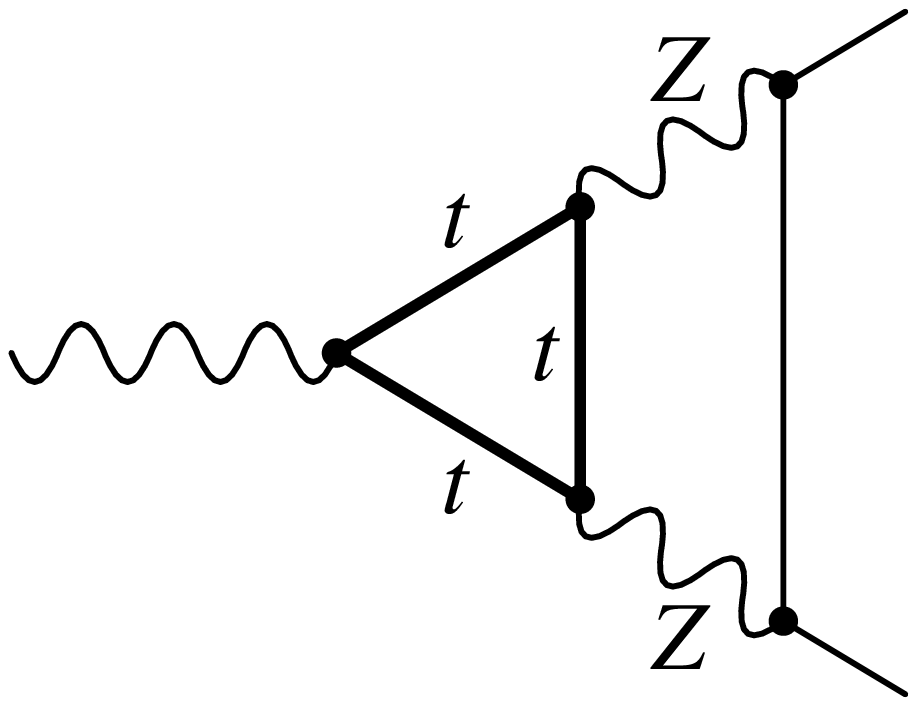, width=4cm, bb=201 300 470 497}
\end{minipage}
\hfill
\begin{minipage}[b]{9cm}
\caption{Example of a two-loop vertex diagram with a top-quark sub-loop.
\label{fig:zt}}
\end{minipage}

\end{figure}
An example of a typical scalar two-loop vertex diagram is shown in Fig.~\ref{fig:zt}.
The expansion of this diagram reads
\begin{equation}
x \, \frac{\zeta(2)}{3} + x^2 \, \left(\frac{\zeta(2)}{12} - \frac{5}{36} +
	\frac{1}{12} \log x \right) + 
  x^3 \, \left( \frac{\zeta(2)}{45} - \frac{79}{1200} + \frac{1}{20} \log x
  	\right) + \dots
\end{equation}
Numerically this amounts to
\begin{equation}
0.1483 - 0.0081 - 0.0019 + 0.0003 + \dots
\end{equation}
The excellent convergence of this series is typical for all diagrams that only
contain neutral current exchange in the loop. Diagrams involving charged current
exchange converge more slowly, which is an effect of the top-bottom mass
splitting.

For this work, the large-mass expansion is executed up to order $x^{5} =
\MZ^{10}/\mt^{10}$, which yields an overall precision of $10^{-7}$, by far
sufficient for practical purposes. This high accuracy is a substantial
improvement over the previous work in Ref.~\cite{ewmt2}, where only the first
two terms in an expansion for large $\mt$ were calculated. Please note that the
large-mass expansion was only used for the two-loop vertex diagrams. The
two-loop counterterms, which in addition to the mass scales $\MW$, $\MZ$ and
$\mt$ also involve the parameter $\MH$, were evaluated using one-dimensional
integral representations as in Refs.~\cite{intnum, intnum2}. In principle it
would also be possible to compute the counterterms using large-mass expansions.
However, since in general analytical results only exist for two-loop diagrams
with up to two different scales, this would require a simultaneous expansion in
$\mt$ and $\MH$, as in Ref.~\cite{ewmt2,ewmt2a}. In order to obtain a precise
result, the one-dimensional integral representations are more suitable instead.

The contributions with light fermions contain only the scales $\MW$ and $\MZ$
and are therefore functions of only one dimensionless variable $\omega =
\MW^2/\MZ^2$. In this case it is possible to evaluate all contributions
analytically using the differential equation method \cite{diffeq}. The final
result is thus expressed through polylogarithms and generalized polylogarithms.

As a simple example consider the scalar integral in Fig.~\ref{fig:lf1}.
Using integration-by-parts identities \cite{ibp}, the following differential
equation can be derived:
\begin{figure}[tb]
\rule{0mm}{0mm}\hfill
\begin{minipage}[b]{8cm}
\raisebox{14mm}{LF1$(p^2,m^2) =$ \ }
\psfig{figure=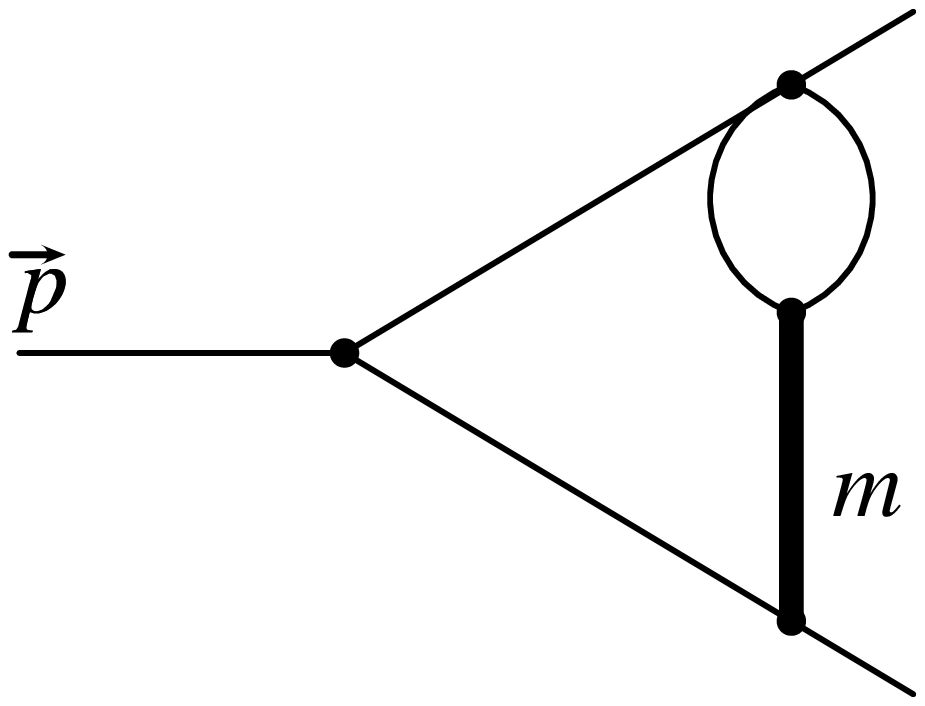, width=4cm}
\end{minipage}
\hfill
\begin{minipage}[b]{7.5cm}
\caption{Example of scalar prototype integral. The thick line is massive with
mass $m$, while the thin lines represent massless propagators.\label{fig:lf1}}
\end{minipage}
\end{figure}
\begin{equation}
\begin{aligned}
p^2 \frac{\rm d}{{\rm d}p^2} \left[
  \raisebox{-6.5mm}{\psfig{figure=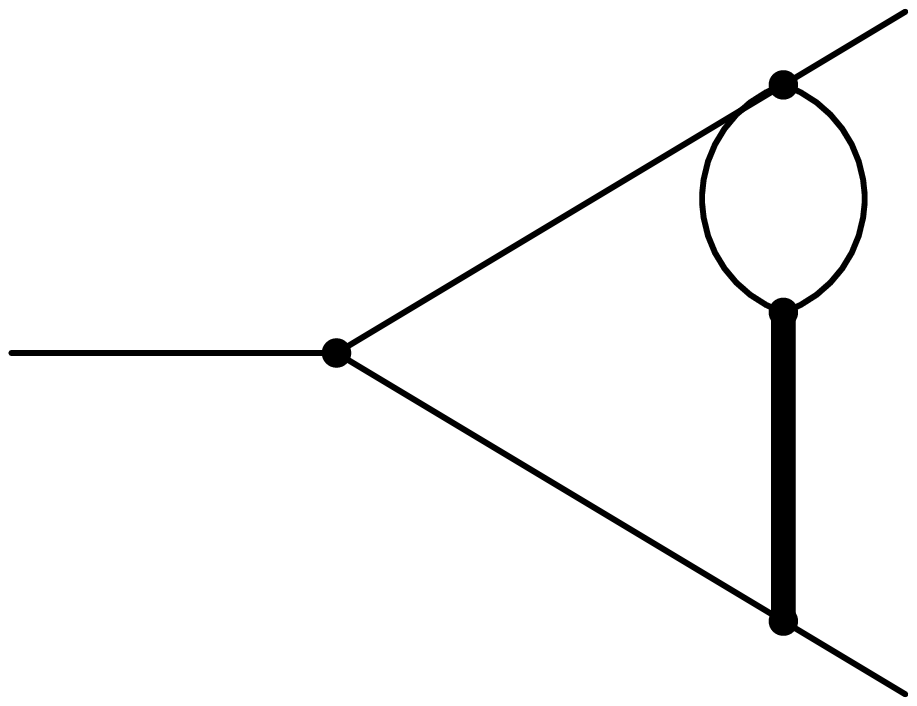, width=2cm}}\;\right]
 = \;
 & \frac{1}{2} \, \frac{p^2}{p^2+m^2} \Biggl( (4-D)(4 + 5 \frac{m^2}{p^2})
  \left[ \raisebox{-6.5mm}{\psfig{figure=v2p1.ps, width=2cm}}\;\right]
  \\
 & + (10 - 3D) \left[ \raisebox{-6.5mm}{\psfig{figure=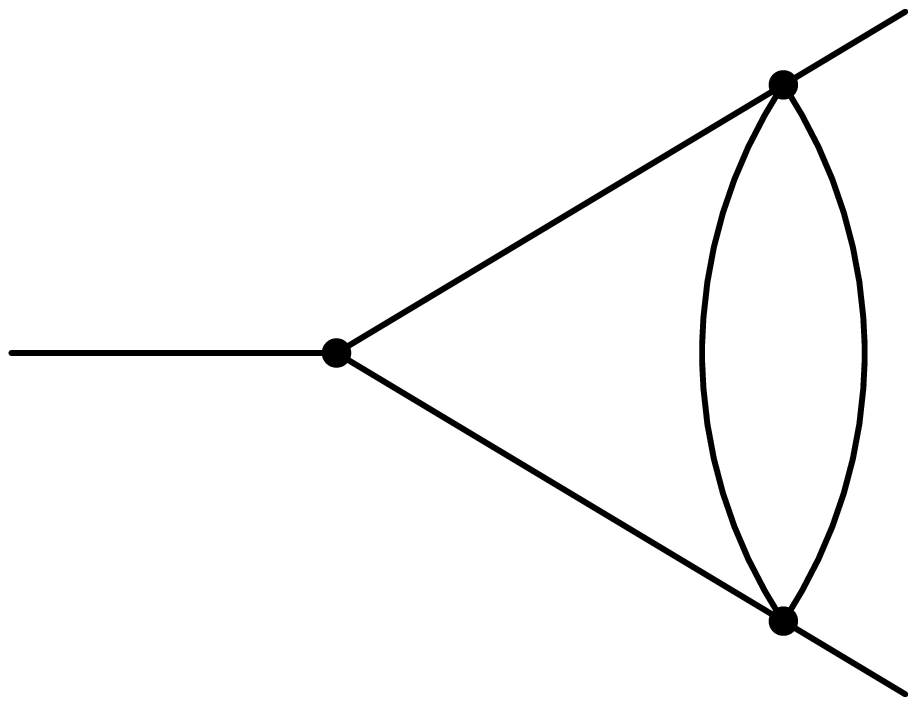, width=2cm}}\;\right]
  - (2-D) \left[ \raisebox{-4mm}{\psfig{figure=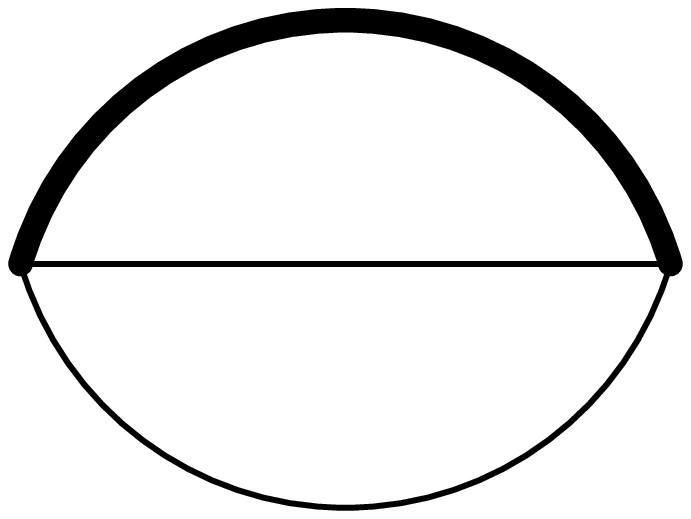, width=1.2cm}}\right]
  \Biggr).
\end{aligned}
\end{equation}
Here the thick lines represent massive propagators with mass $m$ and the thin
lines denote massless propagators. Besides the integral LF1 under study, the
differential equation involves a simpler scalar vertex integral and a vacuum
integral on the right-hand side. Feeding in analytical expression for these
integrals from the literature \cite{davtausk,masslessvertex}, 
the differential equation can be
solved in terms of Nielsen's polylogarithms \cite{nielsen}. 
The finite part of LF1 reads
\begin{equation}
{\rm LF1}(p^2,m^2) = 
\begin{aligned}[t]
&-\li_2(-x) \bigl( -2 + 2 \log (m^2) + 3 \log(-x) + \log(1+x) \bigr)
  + 4 \li_3(-x) - S_{1,2}(-x) \\
&+ \frac{1}{2} \log(1+x) \bigl[ 2 \zeta_2 - \log(-x) \bigl(
  ( -4 + 4 \log(m^2) + 2 \log(-x) + \log(1+x) \bigr) \bigr],
\end{aligned}
\end{equation}
with $x=p^2/m^2$ and 
Nielsen's polylogarithm $S_{1,2}$ defined in
Ref.~\cite{lewin}. The
integral LF1 has also been calculated in Ref.~\cite{Feucht:2003yx}. However,
some of the prototype integrals needed for this project have not been known
before and were computed for the first time in this work. 
All integrals have been checked by different expansions in physical and
unphysical regimes.

Several relevant integrals were also recently computed in Ref.~\cite{agbon}.
However, their results were presented in terms of generalized harmonic
polylogarithms, which in general involve numerical integrations for the
numerical evaluation.

After performing the Dirac and Lorentz algebra for the relevant two-loop vertex
diagrams, the result contains a large number of different scalar integrals with
terms in the numerator that cannot be cancelled against any of the propagators
in the denominator. 
Here it is advantageous to perform an algebraic reduction
to a minimal set of master integrals. 

For the reduction to master integrals, the Laporta algorithm is used
\cite{laporta}. It is based on integration-by-parts \cite{ibp} and Lorentz
identities \cite{lor}, which establish linear relations between scalar
loop integrals. For a sufficiently large set of these relations, the linear
equation system can be solved in order to express the more complicated integrals
with non-trivial numerators in terms of a set of simple master integrals with
unit numerators. This reduction algorithm is implemented in the C++ library
IdSolver \cite{idsolver}, which allows for a fast evaluation of linear 
systems involving several thousand equations.

\begin{figure}[tb]
\begin{center}
\epsfig{figure=v2p0.ps, width=2.5cm}
\epsfig{figure=v2p1.ps, width=2.5cm}
\epsfig{figure=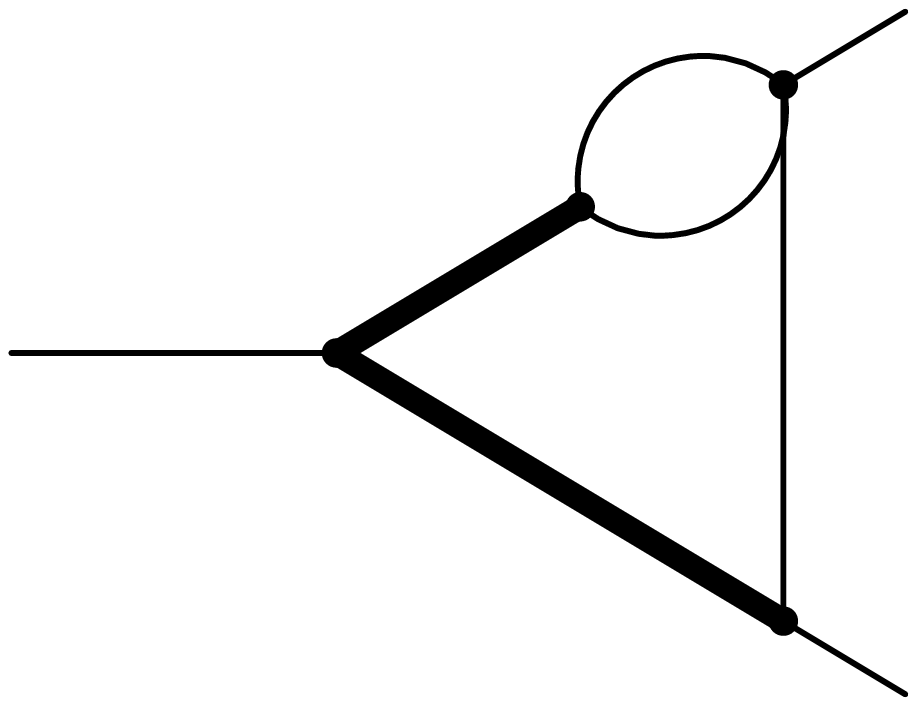, width=2.5cm}
\epsfig{figure=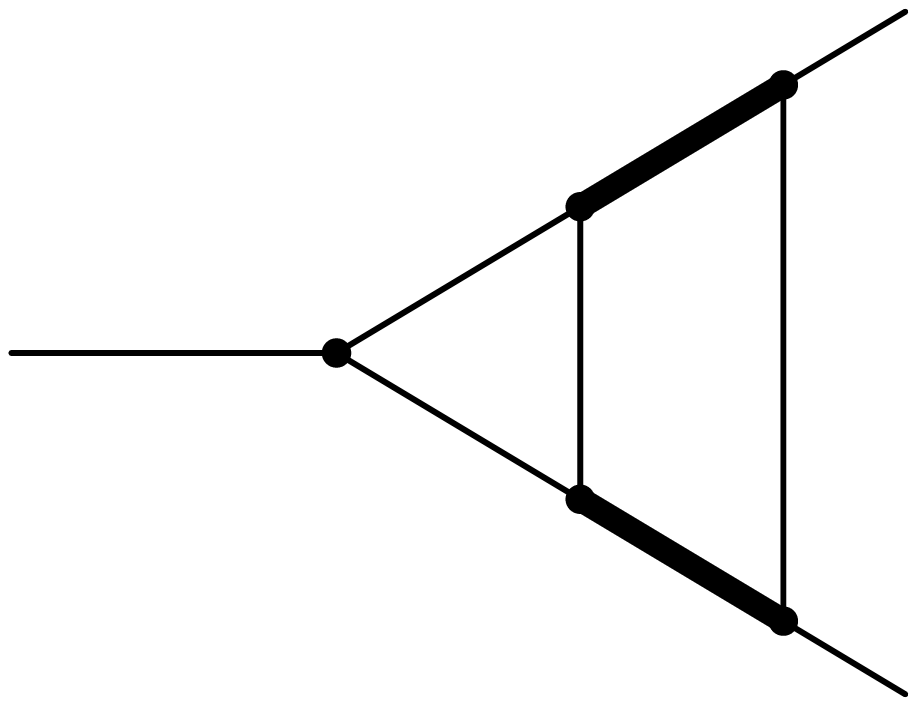, width=2.5cm}
\epsfig{figure=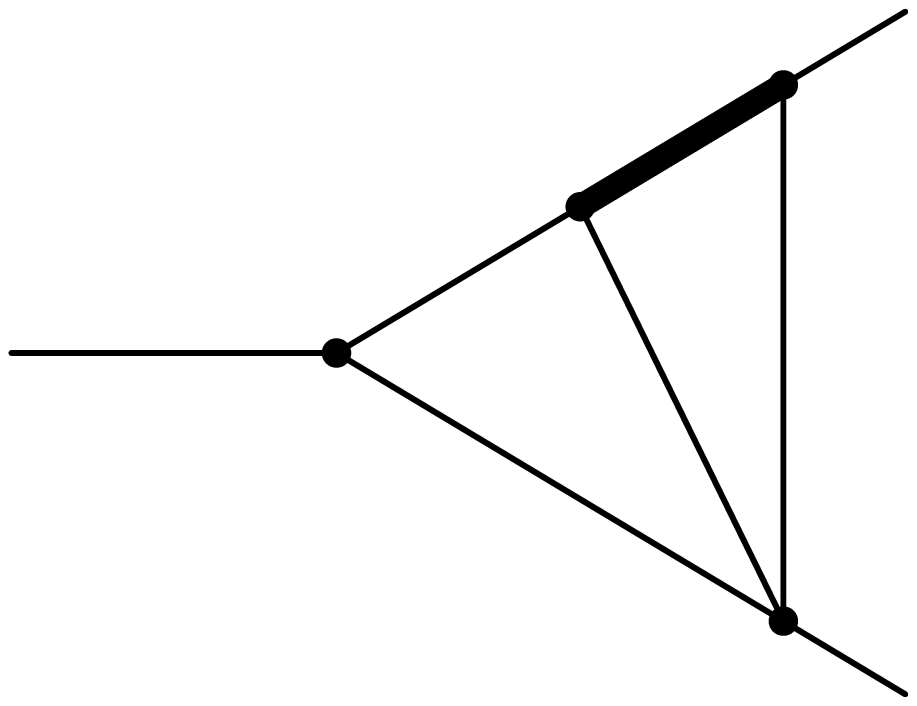, width=2.5cm}
\epsfig{figure=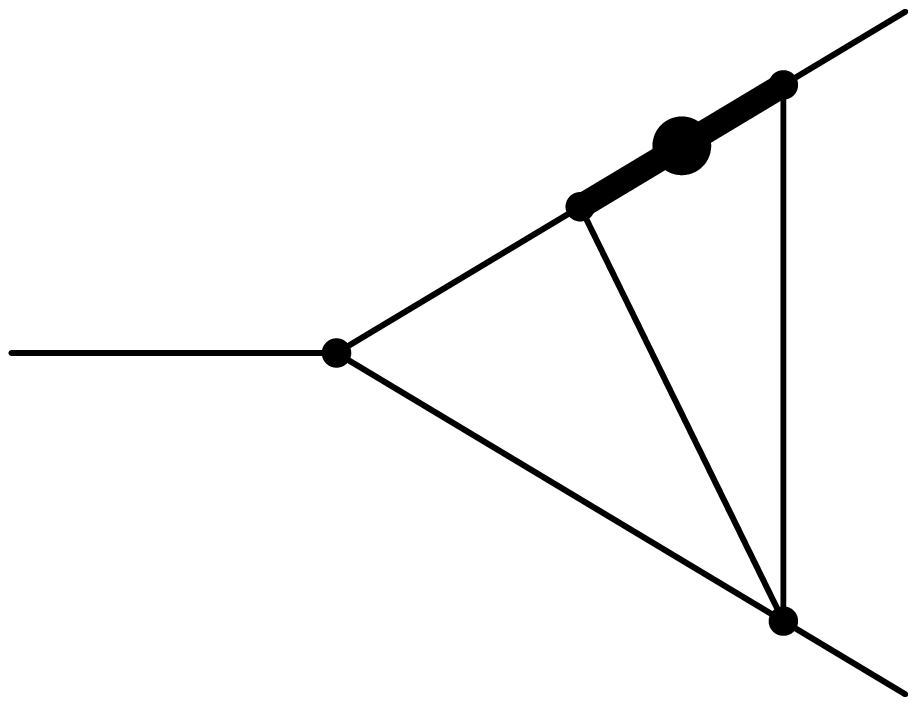, width=2.5cm}
\end{center}
\vspace{-2ex}
\caption{Scalar master integrals for diagrams with a light fermion loop. 
Thick lines indicate massive gauge boson propagators, 
while thin lines correspond to light fermions of photons, which are taken 
massless. The dot in the last diagram indicates that this propagator appears
two times.
\label{fig:masters}}
\end{figure}

The set of master integrals that appear within this calculation for the light
fermion contributions is summarized in Fig.~\ref{fig:masters}. 
Analytical
expressions were found by the differential equation method for all but the
fourth topology in Fig.~\ref{fig:masters}, which was evaluated numerically.

\subsection{Semi-numerical integrations}
\label{sec:disp}

\begin{figure}
\centering
\begin{tabular}{c@{\hspace{2cm}}c}
\psfig{figure=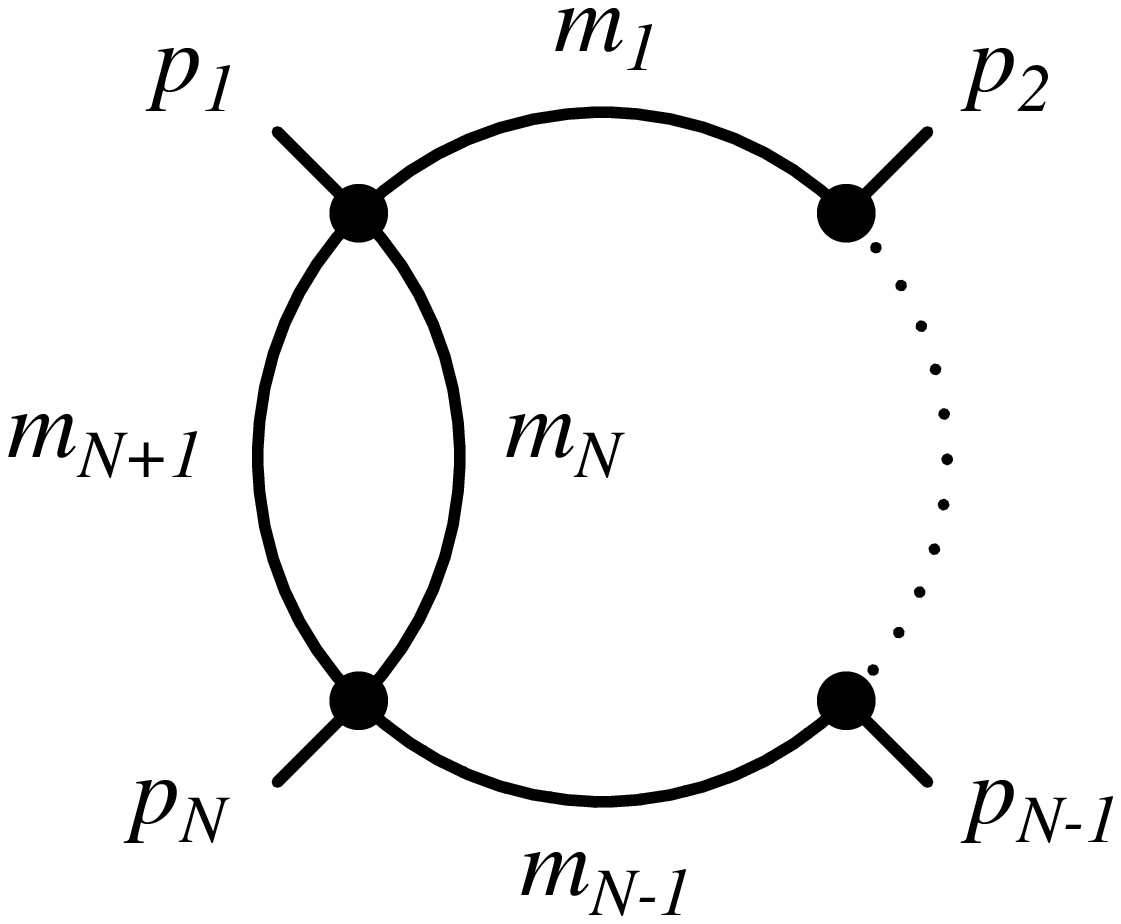, width=3.7cm, bb=103 270 488 542} &
\psfig{figure=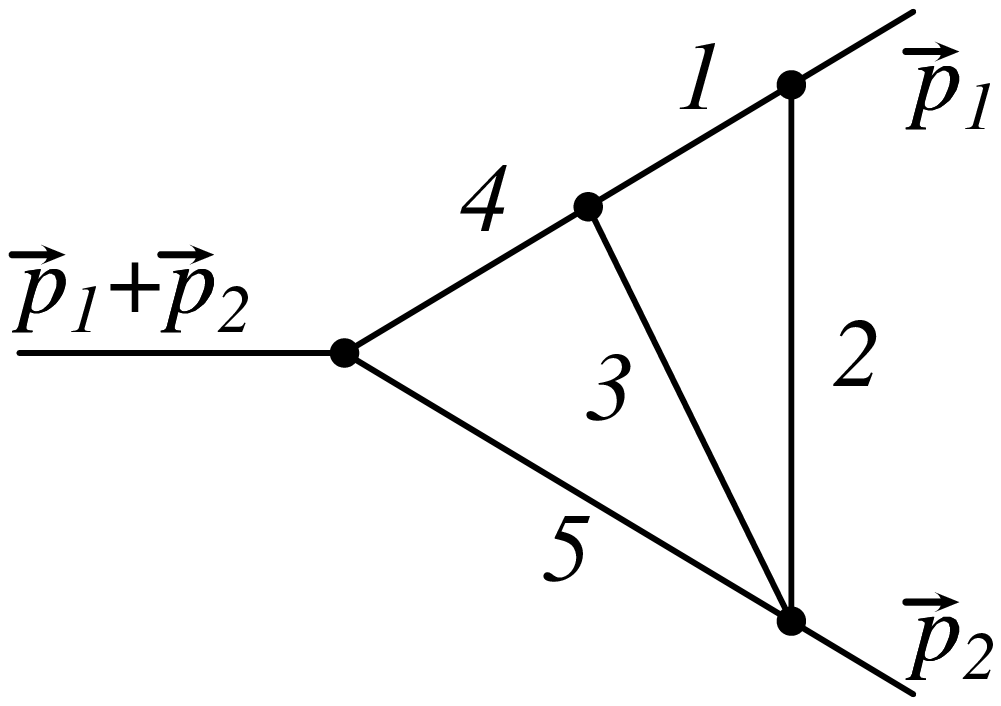, width=3cm}\raisebox{1.15cm}{$\!\!\!\to$} \
\psfig{figure=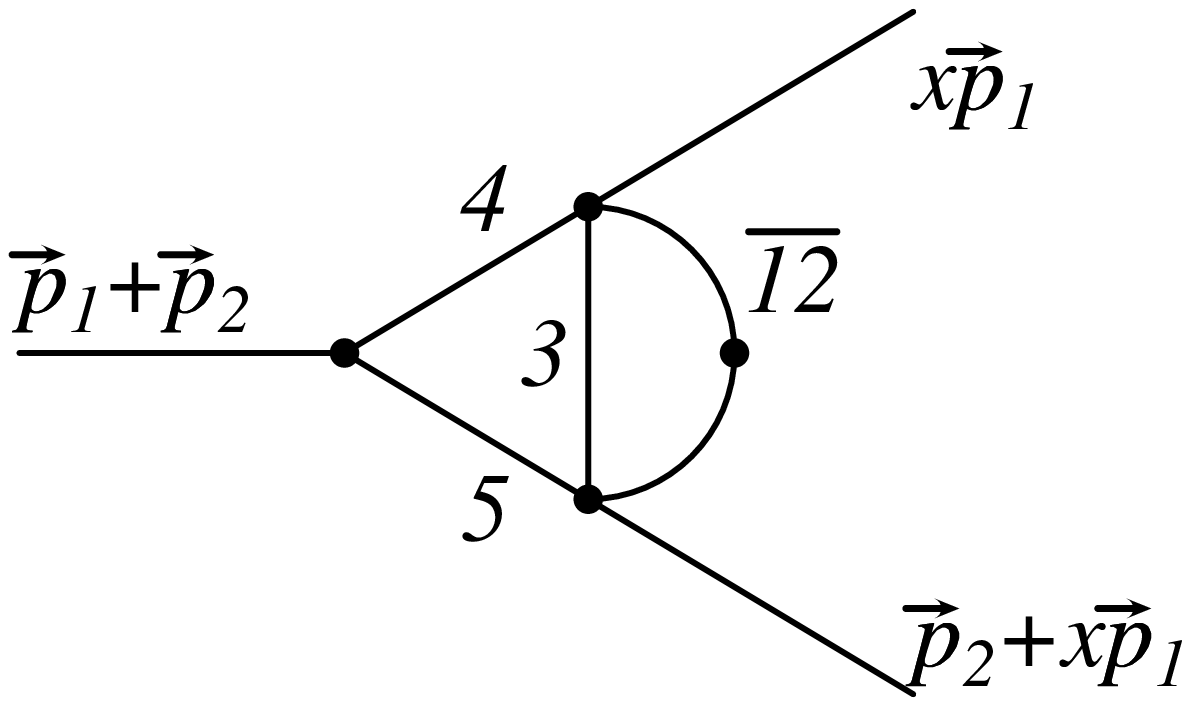, width=4cm} \\
 (a) & (b) \\
\end{tabular}
\caption{(a) General representation of a two-loop scalar diagram with
self-energy sub-loop. (b) Reduction of triangle sub-loop to self-energy sub-loop
by means of Feynman parameters.}
\label{fig:disp}
\end{figure}

The second method employs numerical
integrations for the master integrals. This technique is based on a dispersion
representation of the one-loop self-energy function $B_0$,
\begin{align}
B_0(p^2,m_1^2,m_2^2) &= \int_{(m_1+m_2)^2}^\infty {\rm d}s \,
  \frac{\Delta B_0(s,m_1^2,m_2^2)}{s - p^2}, \\
\Delta B_0(s,m_1^2,m_2^2) &= (4\pi\mu^2)^{4-D} \,
  \frac{\Gamma(D/2-1)}{\Gamma(D-2)} \, \frac{\lambda^{(D-3)/2}(s,m_1^2,m_2^2)}%
  {s^{D/2-1}},
\end{align}
where $D$ is the space-time dimension and $\lambda(a,b,c) = (a-b-c)^2 - 4bc$.
Using this relation, any scalar two-loop integral $T$ with a self-energy
sub-loop as in Fig.~\ref{fig:disp}~(a)
can be expressed as \cite{intnum}
\begin{equation}
\begin{aligned}
T_{N+1}(p_i;m_i^2) = - &\int_{s_0}^\infty {\rm d}s \;
  \Delta B_0(s,m_N^2,m_{N+1}^2) \\
  \times &\int {\rm d}^4 q  \,
  \frac{1}{q^2-s} \,
  \frac{1}{(q+p_1)^2 - m_1^2} \cdots \frac{1}{(q+p_1+\dots+p_{N-1})^2 -
  m_{N-1}^2}.
\end{aligned}
\end{equation}
Here the integral in the second line is a $N$-point one-loop
function, and the integration over $s$ is performed numerically. While in
principle it is also possible to introduce dispersion relations for triangle
sub-loops \cite{intnum2,xloop}, 
it is technically easier to reduce them to self-energy sub-loops by
introducing Feynman parameters \cite{feynpar},
\begin{equation}
\begin{gathered}
\ [(q+p_1)^2-m_1^2]^{-1} \; [(q+p_2)^2-m_2^2]^{-1} = \int_0^1 {\rm d}x
  \; [(q+\bar{p})^2 - \overline{m}^2]^{-2} \\
\bar{p} = x\,p_1 + (1-x)p_2, \qquad
\overline{m}^2 = x \, m_1^2 + (1-x) m_2^2 - x(1-x)(p_1-p_2)^2.
\end{gathered}
\end{equation}
This is indicated diagrammatically in Fig.~\ref{fig:disp}~(b). The integration
over the Feynman parameters is also performed numerically. As a result, all
master integrals for the vertex topologies can be evaluated by at most 3-dim.
numerical integrations. 

The basic scalar two-loop integrals might contain UV- and IR-divergencies. These
need to be subtracted before the numerical integration can be carried out. An
elegant method to remove the divergencies is by subtracting a term from the
integrand that can be integrated analytically. This can be illustrated by the
subtraction of UV divergencies in the following example:
\begin{equation}
\raisebox{-9mm}{\psfig{figure=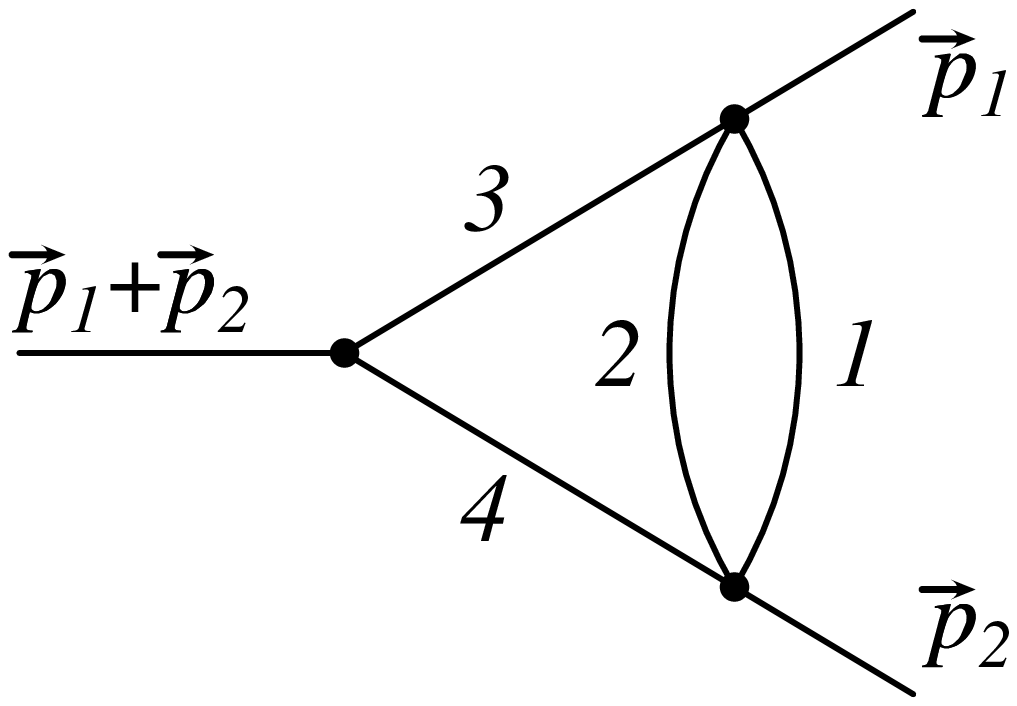, width=2.8cm}} = \;\,
\raisebox{-7mm}{\psfig{figure=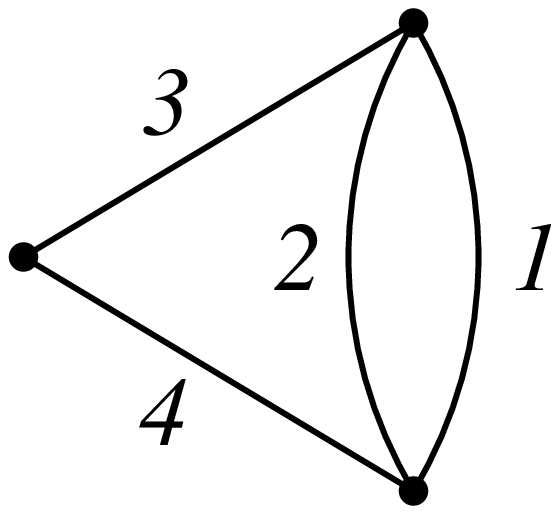, width=1.4cm}} \,\;+\;\;
\raisebox{-9mm}{\psfig{figure=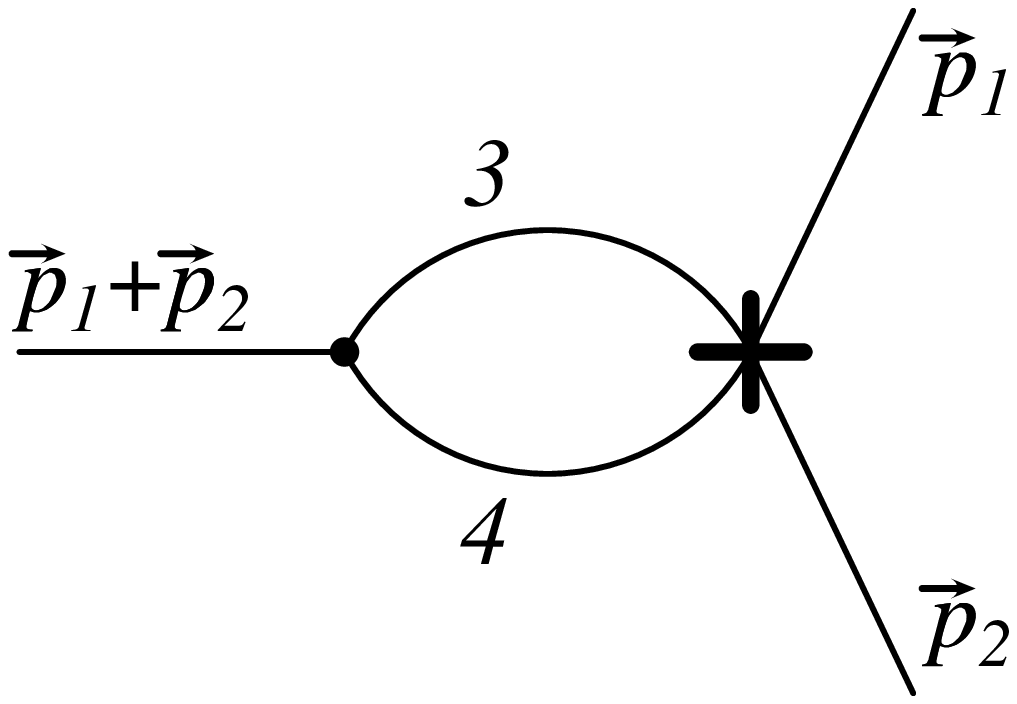, width=2.8cm}} \!-\;\;
\raisebox{-7.5mm}{\psfig{figure=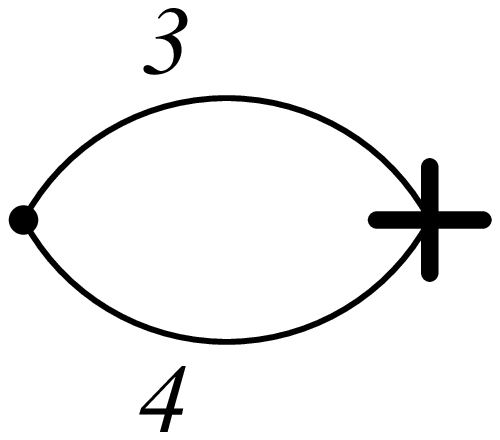, width=1.15cm}} \;\;+\;
\left[ 
\raisebox{-9mm}{\psfig{figure=Tv0.ps, width=2.8cm}} 
\right]_{\rm finite}.
\end{equation}
The UV divergent part of the two-loop vertex diagram can be identified by the
sum of the same diagram with zero external momenta and the contribution from
sub-loop renormalization. The first term corresponds to a two-loop vacuum
diagram for which analytical formulae are available in the literature
\cite{davtausk}, while the second and third terms are products of one-loop functions,
\begin{align}
\raisebox{-9mm}{\psfig{figure=Tvx.ps, width=2.8cm}} &=
B_0 \left( (p_1+p_2)^2, m_3^2, m_4^2 \right) \times 
B_0 \left( m_4^2, m_1^2, m_2^2 \right), \\
\raisebox{-7.5mm}{\psfig{figure=Tvx0.ps, width=1.15cm}} \qquad\, &=
B_0 \left( 0, m_3^2, m_4^2 \right) \times 
B_0 \left( m_4^2, m_1^2, m_2^2 \right).
\end{align}
Here the momentum scale $m_4^2$ for the sub-loop counterterm was chosen to be able to 
handle the case $0 = m_1 = m_2 \neq m_4$.
Subtracting these terms in the integrand of the two-loop vertex
integral results in a finite contribution, that can be integrated numerically,
\begin{equation}
\begin{aligned}
\left[ 
\raisebox{-9mm}{\psfig{figure=Tv0.ps, width=2.8cm}} 
\right]_{\rm finite} \!=\,
&- \int_{(m_1+m_2)^2}^\infty {\rm d} s \;
  \Delta B_0(s,m_1^2,m_2^2) \\[-2ex]
 &\times \biggl[C_0\left((p_1+p_2)^2, p_1^2, p_2^2, m_3^2, m_4^2, s \right) -
 	       C_0\left(0, 0, 0, m_3^2, m_4^2, s \right) \\
   &\qquad + \frac{1}{s-m_4^2} \Bigl[ B_0\left((p_1+p_2)^2, m_3^2, m_4^2\right) -
   			    B_0\left(0, m_3^2, m_4^2\right) \Bigr]
	\biggr].
\end{aligned}
\end{equation}
For all other two-loop vertex master integrals, the divergent parts can be
removed in a similar fashion.

As before, the reduction of integrals with
irreducible numerators to a small set of master integrals is accomplished by
using integration-by-parts and Lorentz-invariance identities, which were
implemented in an independent realization of the Laporta algorithm within
Mathematica.

\subsection{Diagrams with fermion loop triangles and treatment of \boldmath 
$\gamma_5$} 
\label{sc:gamma5}

Diagrams with a fermion triangle sub-loop pose a special problem in conjunction
with the use of dimensional regularization. The fermion triangle loop involves
terms like 
\begin{equation}
{\rm Tr}(\gamma^\alpha\gamma^\beta\gamma^\gamma\gamma^\delta\gamma_5)
 = 4 i \, \epsilon^{\alpha\beta\gamma\delta},
 \label{eq:trg5}
\end{equation}
which cannot be extended to $D$ dimensions simultaneously with the
anti-commutation rule $\{\gamma_\mu, \gamma_5\} = 0$. However, renormalizability
of the Standard Model demands that terms originating from expressions like
eq.~\eqref{eq:trg5} are always UV-finite in any two-loop diagram. As a
consequence, the diagrams with a fermion triangle loop can be treated in two
steps \cite{muon}: 
First the complete diagrams are calculated using naive dimensional 
regularization with anti-commuting $\gamma_5$, where the trace in
eq.~\eqref{eq:trg5} is zero. The finite contributions resulting in epsilon
tensors are computed independently in four dimensions, and finally the two
contributions are added.

\begin{figure}
\centering
\rule{0mm}{0mm}\vspace{-1.2ex}
\begin{tabular}{c@{\hspace{2cm}}c}
\psfig{figure=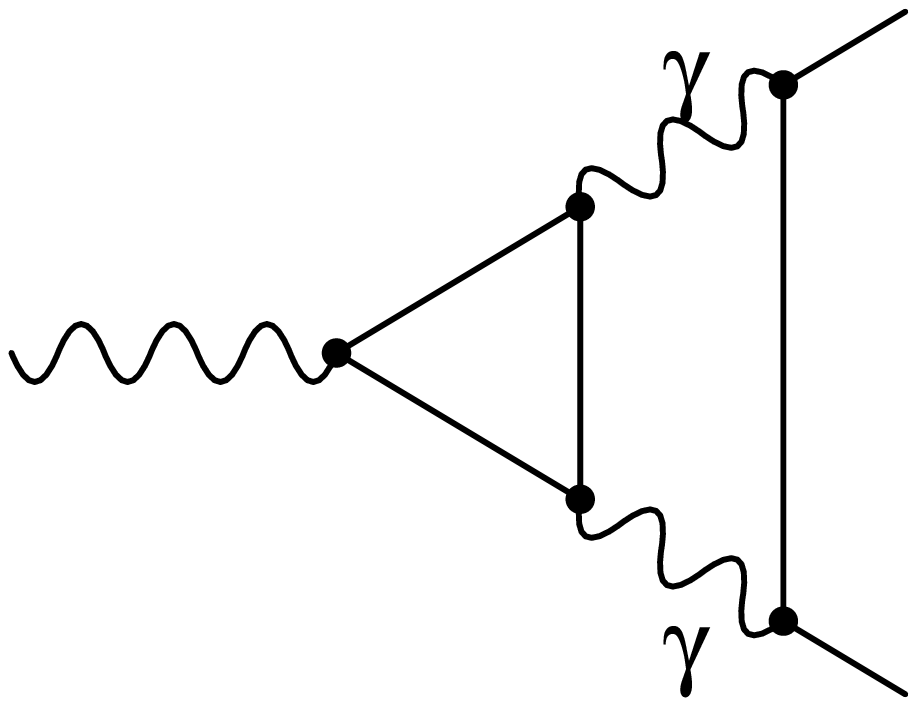, width=4cm} &
\psfig{figure=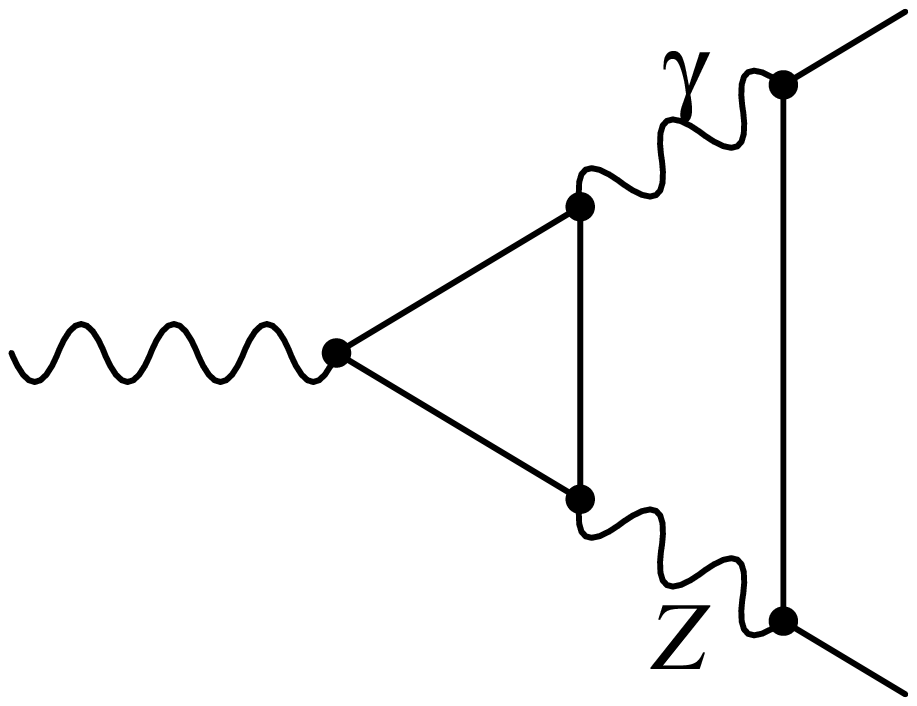, width=4cm} \\[-3ex]
 (a) & (b) \\
\end{tabular}
\caption{Diagrams with fermion triangle sub-loops and soft-collinear 
divergencies.}
\label{fig:trprob}
\end{figure}

An additional complication arises from diagrams with internal photon lines and
massless external fermions, Fig.~\ref{fig:trprob},
which could give rise to soft-collinear divergencies. While these soft and 
collinear divergencies are spurious singularities, thus dropping out in the 
total result, they result in inconsistencies if dimensional regularization 
is used. In this case the contributions involving epsilon tensors from the 
fermion triangle cannot be treated consistently in four dimensions 
anymore.

In this work, the soft and collinear divergencies in these diagrams were 
instead regulated with a photon mass. In the complete result, the limit of 
zero photon mass was taken by means of an expansion, involving a careful 
treatment in the mixed Sudakov/threshold regime. The result for the diagrams
with two photons has been checked against Ref.~\cite{Kniehl:1989qu}.

\subsection{Checks}

The master integrals have been checked with published results where applicable
\cite{Feucht:2003yx, agbon}. Some master integrals were tested by means of
Mellin-Barnes representations, see also \cite{MB,MB2,MB3}, and with a
low-momentum expansion. In addition, complete diagrams were tested with a
low-momentum expansion. In the comparison of the two methods explained in the
previous sections, complete agreement was found.


\section{Calculation of bosonic two-loop vertex diagrams}
\label{sc:2bos}

As explained in the previous chapter, the calculation of the bosonic two-loop
corrections fall into two categories, the bare vertex diagrams and the on-shell
renormalization terms. The computation of the renormalization counterterms has
been established previously \cite{muon2,AC2}, whereas the calculation of the
vertex diagrams will be addressed here. In our case, this involves massive
two-loop three-point function with one massive external leg and up to three
different mass scales.

Contrary to the fermionic corrections, the bosonic diagrams do not depend on the
top quark. On the other hand, there is a dependence on the Higgs boson mass,
which is not a fixed parameter and can assume a broad range of values.
Due to complexity of the problem with several hundred diagrams and many more
different algebraic integral structures, the calculation cannot be performed in
a straightforward way with any known computational method. Here the task is
approached by using an expansion in the various parameters in order to
obtain a result expressed through single scale integrals, which have to be
evaluated numerically in a final step.

In a first step, we apply an expansion in the difference of the
masses of the $W$ and $Z$ bosons, where the expansion parameter is
just $\sw^2$. Since there are diagrams where there is a threshold
when $\MW = \MZ$, the appearance of divergences at higher orders in
the expansion is inevitable.  In this case, we apply the method of
expansions by regions, see \cite{smirnov}. In this approach, one analyzes the momentum
regions which can contribute to the integral and expands the integrand in each
region with a different expansion parameter.
The two regions that
contribute to the result come from the  {\it ultrasoft momenta},
$q_{1,2} \sim \sw^2 \MZ$,  and {\it hard momenta}, $q_{1,2} \sim
\MZ$, where $q_{1,2}$ are the loop momenta.  
Then the reduction to the set
of master integrals proceeds with Integration-By-Parts identities
\cite{ibp} solved with the Laporta algorithm \cite{laporta} as
implemented in the {\it IdSolver} library \cite{idsolver}.

The $\MH$ dependence is treated in two regimes. 
For low values of $\MH$ an expansion
in the mass difference between $\MH$ and $\MZ$ is used, with the expansion
parameter defined to be
\begin{equation}  
    s_{\rm H}^2=1-\frac{\MH^2}{\MZ^2},
\end{equation}  
where this time no non-trivial thresholds are encountered. 
It is found that a good precision is achieved by performing the expansion to the
sixth order in $\sw^2$ and $s_{\rm H}^2$.
The second regime is for large values of $\MH \gg \MZ$, where a large mass
expansion \cite{smirnov} is used.

The resulting single scale master integrals are treated with various methods,
usually with two or three different ones for test purposes. Most integrals can
be obtained with numerical integrations based on dispersion relations as described
in section \ref{sec:disp}. The advantage of this method is that with reasonable
investment of computer time, it can be pushed to high precision, which is
required since large numerical cancellations are observed between individual
integrals. Diagrams of simpler topologies can also be
evaluated with differential equations \cite{Kotikov:1991hm,Remiddi:1997ny} and
large mass expansions. For more complicated topologies, Mellin-Barnes
representations are employed, using the {\sc MB} package \cite{MB}, see also
\cite{MB2,MB3}. After simplification, the Mellin-Barnes representations can be
evaluated by numerical integrations or infinite series. In principle, this
method could be used for all scalar integrals, however, depending on the mass
configuration, the integration and/or the series evaluation does not converge.
The convergence behavior can be improved by rotating the integration contours
into the complex plane, but this also solves the problem only in a few cases.
Whenever possible the results were cross-checked with sector
decomposition~\cite{sectors}.

The reduction to master integrals can occasionally can produce spurious $1/(D-4)$
poles in the coefficients of some master integrals. In principle, this problem
can be avoided by choosing an appropriate basis of master integrals, at
the expense, however, that some of these integrals are more complicated. Here, on the other hand, a basis was chosen that introduces only relatively few spurious
poles, but in front of simple integrals. Since it is advantageous to check the
cancellation of divergencies exactly, it was thus necessary to evaluate the
finite pieces of some master integrals analytically. These integrals are
presented in Ref.~\cite{radcor05}.

As a final algebraic check of the whole procedure, the cancellation of the gauge
parameter dependence in a general covariant $R_\xi$ gauge was verified. Due to
the enormous complexity of the intermediate expressions, this test was only
possible for the first orders in the expansion, but nevertheless allowed a
non-trivial cross-check between different diagram topologies.


\section{Numerical Results}
\label{sc:results}

In order to arrive at a precise prediction for the effective weak mixing angle,
the electroweak corrections of one- and two-loop order are combined with one-
and two-loop QCD corrections \cite{qcd2,qcd3}, and leading three-loop
corrections of order ${\cal O}(\GF^3 \mt^6)$ and ${\cal O}(\GF^2 \alps \mt^4)$
\cite{mt6}. Other higher-order corrections to the rho parameter of order ${\cal
O}(\GF^3 \MH^4)$   \cite{radja} and ${\mathcal O}(G_\mu m_t^2 \alps^3)$
\cite{qcd4} are very  small (for $\MH < 1$ TeV) and thus not included in the
numerical analysis. The result is expressed as a perturbative expansion in
$\alpha$, not $G_\mu$.
Instead,
all higher-order reducible contributions, that arise from terms proportional to
$\De\al$ and $\De\rho$, are included explicity at the given loop order in the
computation. A finite $b$ quark mass was retained in the ${\cal O}(\al)$ and
${\cal O}(\al\alps)$ contributions, but neglected in all higher-order terms.

In Tab.~\ref{tab:kappa}, the effects of the various loop contributions on the
vertex form factor $\De\kappa$ are shown for the input parameters in
Tab.~\ref{tab:input}.
$\De\al$ is defined as the real part of the shift of the photon vacuum
polarization function $\Pi(q^2)$ between $q^2=0$ and $q^2=\MZ^2$ that
stems from light fermions,
\begin{equation}
\De\al = \re \bigl\{ \Pi_{\rm lf}(0)-\Pi_{\rm lf}(\MZ^2) \bigr\},
\qquad
\Pi(q^2) = \Pi_{\rm lf}(q^2) + \Pi_{\rm rest}(q^2).
\end{equation}
It is important to note that the experimental values for
the $W$ and $Z$ boson masses in Tab.~\ref{tab:input} correspond to a
Breit-Wigner parametrization with a running width, that have to be translated to
the pole mass scheme used in the loop calculations~\cite{muon}. In effect, this
translation results in a downward shift \cite{riemann} of $\MZ$ by 34 MeV and
$\MW$ by 28 MeV, respectively.
\begin{table}[tb]
\begin{center}
\begin{tabular}{ll}
\hline
      Input parameter & Value\\ 
      \hline 
      $\MW$ & $80.404 \pm 0.030 \gev$ \\
      $\MZ$ & $91.1876 \pm 0.0021 \gev$ \\
      $\Gamma_Z$ & $2.4952 \gev$ \\
      $m_t$ & $172.5 \pm 2.3 \gev$ \\
      $m_b$ & $4.85 {\rm \; GeV}$ \\
      $\Delta\alpha(\MZ^2)$ & $0.05907 \pm 0.00036$ \\
      $\alpha_s(\MZ)$ & $0.119 \pm 0.002$ \\
      $G_\mu$ & $1.16637 \times 10^{-5} \gev^{-2}$ \\
\hline
\end{tabular}
\end{center}
\vspace{-1em}
\caption{Experimental input parameters used in the numerical evaluation; from
Refs.~\cite{lepewwg,Eidelman:2004wy}.
\label{tab:input}}
\end{table}
\begin{table}[tb]
\begin{center}
\begin{tabular}{lrrrrrrrr}
\hline
$\MH$ & ${\cal O}(\al)$ & ${\cal O}(\al^2)_{\rm ferm}$ & 
${\cal O}(\al^2)_{\rm bos}$ & 
${\cal O}(\al\alps)$ &
${\cal O}(\al\alps^2)$ &
${\cal O}(\al^2\alps\mt^4)$ & 
${\cal O}(\al^3\mt^6)$ & 
red. \\ 
\rule{0mm}{0mm}[GeV] \ & 
[$10^{-4}$] \\
\hline 
100 & 413.33 & 1.07 & -0.74 & -35.58 & -7.25 & 1.15 & 0.14 & 0.69 \\
200 & 394.02 &-0.32 & -0.47 & -35.58 & -7.25 & 1.90 & 0.07 & 0.70 \\
600 & 354.06 &-2.89 &  0.17 & -35.58 & -7.25 & 3.70 & 0.08 & 0.72 \\
1000& 333.16 &-2.61 &  1.11 & -35.58 & -7.25 & 4.53 & 0.91 & 0.72 \\
\hline
\end{tabular}
\end{center}
\vspace{-1em}
\caption{Loop contributions to $\Delta\kappa$ with fixed $\MW$ as input
parameter as a function of the Higgs mass $\MH$. Here "red." corresponds to
reducible three-loop contributions stemming from $\De\al$ and $\De\rho$.
\label{tab:kappa}}
\end{table}

\begin{figure}[tb]
\begin{center}
\epsfig{figure=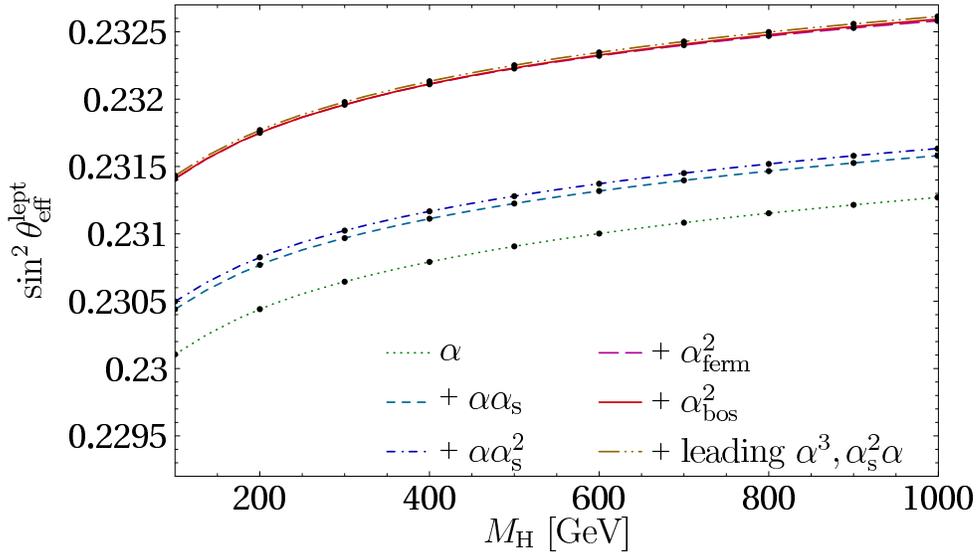, width=13cm, bb=20 489 363 682}
\end{center}
\vspace{-1.5em}
\caption{Contribution of several orders of radiative corrections to the
effective leptonic weak mixing angle \SinEff as a function of the Higgs mass
\MH. The tree-level value is not shown.
\label{fig:sweff}}
\end{figure}

As evident from the table, the fermionic and bosonic contributions to
$\De\kappa$ are of the same magnitude. This changes, however, when expressing
the result through the Fermi constant $G_\mu$ as input parameter. For this, the
corresponding loop corrections, $\De r$, to the $W$ boson mass need to be 
incorporated,
\begin{equation}
\MW^2 \left(1 - \frac{\MW^2}{\MZ^2}\right) =
\frac{\pi \al}{\sqrt{2} \GF} \left(1 + \De r\right).
\label{eq:delr}
\end{equation}
The inclusion of the corrections to $\MW$ lead to an enhancement of the
fermionic two-loop corrections to \SinEff, but to a partial cancellation
between the bosonic two-loop corrections in $\De\kappa$ and $\De r$.
The effect of the different loop orders in \SinEff with $G_\mu$ as input
parameter is summarized in Fig.~\ref{fig:sweff}.
The figure shows that the contribution from the fermionic two-loop corrections
amount to roughly $\sim 10^{-3}$, while the resulting effect of the bosonic two-loop
corrections is about or less than $\sim 10^{-5}$, so that the two curves for 
${\cal O}(\al+\al\alps+\al\alps^2+\al^2_{\rm ferm})$ and 
${\cal O}(\al+\al\alps+\al\alps^2+\al^2_{\rm ferm}+\al^2_{\rm bos})$ practically overlap.

For the analysis in the following sections, the new full result always includes
terms of the orders $\al$, $\al^2$, $\al\alps$, $\al\alps^2$, $\al^2\alps\mt^4$
and $\al^3\mt^6$,
\begin{equation}
\sineff\bigr|_{\rm full} =
\sineff\bigr|_{\al+\al^2+\al\alps+\al\alps^2+\al^2\alps\mt^4+\al^3\mt^6}.
\label{eq:swfull}
\end{equation}

\subsection{Comparison with previous results}

The most precise previous result for the two-loop electroweak corrections to
\SinEff was obtained from the calculation of the next-to-leading term ${\cal O}(\GF^2 
\Mt^2\MZ^2)$ in an expansion for large values of the top-quark mass $\mt$
\cite{ewmt2}. The impact of the new result, as defined in eq.~\eqref{eq:swfull}, 
is shown in Tab.~\ref{tab:compare}~(a) by comparing with the previous result as
in the fitting formula in
Ref.~\cite{mt2fit} and in the implementation of the program {\sc Zfitter 5.10}
(and later versions) \cite{zfitter}.
\begin{table}[tb]
\begin{center}
\begin{tabular}[t]{lcc}
{\bf (a)} \\
\hline
$\MH$ & $\bigl[\Delta\sineff\bigr]_{\rm ZFITTER}$ & 
 $\bigl[\Delta\sineff\bigr]_{\!\!\!\mbox{\scriptsize\cite{mt2fit}}}$ \\
\rule{0mm}{0mm}[GeV] \ & 
[$10^{-4}$] & [$10^{-4}$] \\
\hline 
100 & -0.45 & -0.40 \\
200 & -0.69 & -0.72 \\
600 & -1.17 & -0.94 \\
1000& -1.60 & -1.28 \\
\hline
\end{tabular}
\hspace{1ex}
\begin{tabular}[t]{lrrl}
{\bf (b)} \\
\hline
$\mt,\MH$ & $\Delta[\mt^4]$ & $\Delta[\mt^2]$ &
 $\Delta[\mt^{-4}]$ \\
\rule{0mm}{0mm}[GeV] \ & 
 \\
\hline 
175,400 & 20\% & 4.3\% & 0.02\% \\
800,1800 & 5\% & 1.9\% & 0.00002\% \\
\hline
\end{tabular}
\end{center}
\vspace{-1em}
\caption{(a) Difference between the new result of eq.~\eqref{eq:swfull} and the previous
result from Ref.~\cite{ewmt2}, as implemented in {\sc Zfitter} (left column) and
from the fitting formula in Ref.~\cite{mt2fit} (right column).
(b) Convergence of the expansion in $\mt^{-2}$ for the two-loop diagrams with top
propagators. Here $\Delta[\mt^k] =
[\sineff]_{(\al^2\mt^k)}/[\sineff]_{(\al^2\rm exact)}-1$ 
is the relative difference
between the exact and the expanded result at the given order.
\label{tab:compare}}
\end{table}

A more detailed analysis reveals that there are several sources for the
deviations listed in Tab.~\ref{tab:compare}~(a). First of all, there is the
effect of the truncated series expansion in $\mt^{-2}$, which was evaluated only
up to order $\mt^2$ in Ref.~\cite{ewmt2}. In addition, the genuine light-fermion
two-loop contributions were not included in that work. 
Moreover, the
implementation of the correction form factor $\Delta r$ to the $W$ mass 
and the parametrization with $G_\mu$ instead of $\alpha$ in Ref.~\cite{ewmt2} 
introduces 
higher-order terms that can be sizeable. Here it is
important to note that the OSI scheme in Ref.~\cite{ewmt2}, which is the
basis for the implementation of these corrections in ZFITTER,
uses the $\overline{\rm MS}$ definition
for $\Delta\rho$, which is numerically larger than the leading $\mt^2$ term, so
that the resummation effects of $\Delta\rho^{\overline{\rm MS}}$ are rather
large. Finally, {\sc Zfitter} versions before 6.40 use an outdated implementation
of the QCD corrections. Since all these contributions are non-negligible at the
current level of precision, it is interesting to study them separately.

In particular, using the results of section~\ref{sc:largemt} the effect of the
truncated top-mass expansion is shown in Tab.~\ref{tab:compare}~(b)\footnote{As
a by-product of this comparison, we found a typo in Ref.~\cite{ewmt2a}, where a
term $\tfrac{3}{2} \mt^2/(\MZ^2 \sw^2) \log{\cw^2}$ is missing in the
expression for $\MH \gg \mt$.}. It turns out that the expansion converges quite
well for realistic values of $\mt$ and $\MH$. However, the terms beyond the
order $\mt^2$ induce a difference of 4.3\% in the two-loop corrections 
with top-bottom loops,
corresponding to a shift of about $0.2 \times 10^{-4}$ in
\SinEff, which is roughly a quarter of the total difference
reported in Tab.~\ref{tab:compare}~(a). As a cross-check, also the result for
very large values of $\mt$ and $\MH$ are shown in Tab.~\ref{tab:compare}~(b),
to illustrate that in this case the series converges much faster.

\subsection{Error estimate}

While the inclusion of the fermionic two-loop corrections is a substantial
improvement of the prediction of \SinEff\ in the Standard Model, uncertainties
from missing higher order contributions can still be sizeable. Here we try to
give an estimate of the error induced by these unknown contributions. The most
relevant missing higher order contributions are corrections of the order ${\cal
O}(\alpha^2\alps)$ beyond the leading $\mt^4$ term, ${\cal O}(\alpha^3)$ beyond
the leading $\mt^6$ term and ${\cal O}(\alpha \alps^3)$. Since the final
prediction for \SinEff\ is based on \GF\ as input, the loop effects in the both
quantities $\Delta r$ (for the computation of \MW) and $\De\kappa$ (for the $Z
l^+ l^-$ vertex corrections) need to be considered. 

When combining the two form factors, it turns out that there are some
cancellations between the known corrections to \MW\ and the $Z$ vertex. It is
expected that similar cancellations occur when adding an additional QCD loop,
since QCD corrections enter with the same relative sign in the corrections to
$\MW$ and the $Z$ vertex. Since the dominant missing higher order effects
are contributions with an additional QCD loop, it is assumed in the following
that these cancellations are natural and it is justified to study the
theoretical error of both quantities $\Delta r$ and $\De\kappa$ in conjunction.

A simple method to estimate the higher order uncertainties is based on the
assumption that the perturbation series follows roughly a geometric progression.
This presumption implies relations like
\begin{equation}
{\cal O}(\al^2\alps) = \frac{{\cal O}(\al^2)}{{\cal O}(\al)} \, {\cal O}(\al\alps).
\end{equation}
From this one obtains the error estimates in the second column of
Tab.~\ref{tab:error} for the different higher order contributions, which are
given for a range of the Higgs \MH\ mass between 10 GeV and 1000 GeV.
\begin{table}[tb]
\begin{center}
\begin{tabular}{llll}
\hline
 & Geometric progression & Scale dependence & Leading $\mt$ terms\\
\hline
${\cal O}(\alpha^2 \alps)$ beyond leading $\mt^4$ & $3.3 \dots 2.8 \times
10^{-5}$
 & $0.8\dots2.1 \times 10^{-5}$ &
 $1.2\dots4.3 \times 10^{-5}$ \\
${\cal O}(\alpha \alps^3)$ & $1.5 \dots 1.4$ &
 $0.3\dots0.2$ \\
${\cal O}(\alpha^3)$ beyond leading $\mt^6$ & 
$2.5 \dots 3.5$ && $0.3\dots0.8$
\\
\hline \hline
Sum & $4.4 \dots 4.7 \times 10^{-5}$ \\
\hline
\end{tabular}
\end{center}
\vspace{-1em}
\caption{Estimation of the uncertainty from 
different higher order contributions for \SinEff, with the quadratic sum of all
error sources. Where applicable, two or three
different methods for the error estimate have been used.
\label{tab:error}}
\end{table}
To account for possible deviations from the geometric series behavior, an
ad-hoc overall factor $\sqrt{2}$ was included in all error determined via this method.

Alternatively, the error from a higher-order QCD loop can be assessed by
varying the scale of the strong coupling constant $\alps$ or the top-quark mass
\mt\ in the $\overline{\rm MS}$ scheme in the highest available perturbation
order. By varying thus the scale $\mu$ of $m_{\rm t,\overline{MS}}$ in the
${\cal O}(\al^2)$ contributions between $\mt^2/2 < \mu^2 < 2 \mt^2$ one obtains
an error estimate for the ${\cal O}(\al^2\alps)$ contributions between 0.1 and
$3.9 \times 10^{-5}$, depending on the value of $\MH$ for $10 \gev < \MH <
1000 \gev$. Similarly, by varying $\alps(\mu)$ in the ${\cal
O}(\al\alps^2)$ corrections between $\mt^2/2 < \mu^2 < 2 \mt^2$ leads to an
error estimate for the ${\cal O}(\al\alps^3)$ contributions of less than
$10^{-6}$, see Tab.~\ref{tab:error}. 

An independent third estimate of the error of the ${\cal O}(\al^2\alps)$ and
${\cal O}(\al^3)$ contributions can be obtained from the existing leading terms
in the expansion for large top quark mass. Experience from the ${\cal O}(\al^2)$
corrections suggests that for moderate values of $\MH$, the leading $\mt$-term
and the remaining non-leading terms are of similar order. These contributions
are shown in the last column of Tab.~\ref{tab:error}. 

As evident from the table, all methods give results of similar order of
magnitude, while the geometric
progression method tends to lead to the largest error evaluation. The total
estimated error is therefore computed by summing in quadrature the error from
different contributions obtained by this method. It is found to amount to
$\delta_{\rm th}\!\sineff = 4.7 \times 10^{-5}$.

\subsection{Parametrization formulae}
\label{sec:param}

Following Ref.~\cite{sineff}, the numerical results are expressed in
terms of a fitting formula, which reproduces the exact calculation with maximal
and average deviations of $4.5\times10^{-6}$ and $1.2\times 10^{-6}$,
respectively, as long as the input parameters stay within their $2\sigma$
ranges and the Higgs boson mass in the range 10 GeV $\leq M_H \leq$ 1 TeV.
For the sake of comparability with the result of Ref.~\cite{sineff}, the
slightly outdated central values for the experimental input parameters used
there are also kept in the formula
\begin{equation}
\label{eq:formula}
\begin{aligned}
\sinefff = s_0 &+ d_1 L_H + d_2  L_H^2 + d_3  L_H^4 + d_4  (\Delta_H^2 -1) 
 + d_5  \Delta_\alpha \\&+ d_6  \Delta_t + d_7  \Delta_t^2 
 + d_8  \Delta_t  (\Delta_H -1)
 + d_9  \Delta_{\alps} + d_{10} \Delta_Z,
\end{aligned}
\end{equation}
with
\begin{equation}
\begin{aligned}
L_H &= \log\left(\frac{M_H}{100 \gev}\right), &
\Delta_H &= \frac{M_H}{100 \gev}, &
\Delta_\alpha &= \frac{\Delta \alpha}{0.05907}-1, \\
\Delta_t &= \left(\frac{m_t}{178.0 \gev}\right)^2 -1, &
\Delta_{\alps} &= \frac{\alps(\MZ)}{0.117}-1, &
\Delta_Z &= \frac{\MZ}{91.1876 \gev} -1.
\end{aligned}
\end{equation}
The values of the coefficients for the effective leptonic weak mixing angle
$\sineff$ are given in the second column of Tab.~\ref{tab:flav}. This
parametrization includes all relevant known corrections at this time, as in
eq.~\eqref{eq:swfull}.

For some purposes, it is however useful to have a numerical result for the
two-loop electroweak form factors $\De\kappa$ and $\De r$ alone. For
$\De\kappa$, the following parametrization provides a good approximation,
\begin{align}
\label{eq:formkap}
\De\kappa^{(\al^2)} &= \De\alpha \, \De\kappa^{(\al)} +
\De\kappa^{(\al^2)}_{\rm rem}, \\
\De\kappa^{(\al^2)}_{\rm rem} &= 
\begin{aligned}[t]
k_0 &+ k_1 L_H + k_2  L_H^2 + k_3  L_H^4 + k_4  (\Delta_H^2 -1) 
 + k_5  \Delta_t + k_6  \Delta_t^2 
 + k_7  \Delta_t L_H \\&+ k_8 \Delta_W + k_9 \Delta_W \Delta_t
 + k_{10} \Delta_Z,
\end{aligned}
\intertext{with}
\Delta_W &= \frac{\MW}{80.404 \gev} -1.
\end{align}
From a fit to the exact computation, the coefficients are obtained as
\begin{equation}
\begin{aligned}
k_0 &= -0.002711, & k_1 &= -3.12 \times 10^{-5}, & k_2 &= -4.12\times 10^{-5}, &
k_3 &= 5.28\times 10^{-6}, \\
k_4 &= 3.75\times 10^{-6}, & k_5 &= -5.16\times 10^{-3}, & k_6 &= -2.06\times
10^{-3}, & k_7 &= -2.32\times 10^{-4}, \\
k_8 &= -0.0647, & k_9 &= -0.129, & k_{10} &= 0.0712.
\end{aligned}
\end{equation}
This reproduces the exact result for $\De\kappa^{(\al^2)}$ with maximal
deviations of $1.8\times 10^{-5}$ for 10 GeV $\leq
M_H \leq$ 1~TeV and the other input parameters in their $2\sigma$
ranges. This error in $\De\kappa$ corresponds to an error of $4\times 10^{-6}$
for $\sineff$.
Since the experimental values for the top quark mass and the $W$-boson mass
might change substantially with future updates of measurements from the
Tevatron and the LHC, it is useful to see how well the fitting formula
works for larger ranges of these two parameters.
If the top quark mass and the $W$-boson mass vary within $4\sigma$
of their current experimental uncertainty, the formula eq.~\eqref{eq:formkap} is
still accurate to $3.6\times 10^{-5}$, corresponding to an error of $8\times
10^{-6}$ for $\sineff$.

Similarly, for $\Delta r$, the numerical result can be cast into the form
\begin{align}
\label{eq:formdr}
\De r^{(\al^2)} &= (\De\alpha)^2 +  2\De\alpha \, \De r^{(\al)} + \De
r^{(\al^2)}_{\rm rem},
\\
\De r^{(\al^2)}_{\rm rem} &= 
\begin{aligned}[t]
r_0 &+ r_1 L_H + r_2  L_H^2 + r_3  L_H^4 + r_4  (\Delta_H^2 -1) 
 + r_5  \Delta_t + r_6  \Delta_t^2 
 + r_7  \Delta_t L_H \\&+ r_8 \Delta_W + r_9 \Delta_W \Delta_t
 + r_{10} \Delta_Z,
\end{aligned}
\end{align}
where
\begin{equation}
\begin{aligned}
r_0 &= 0.003354, & r_1 &= -2.09\times 10^{-4}, & r_2 &= 2.54\times 10^{-5}, &
r_3 &= -7.85\times 10^{-6}, \\
r_4 &= -2.33\times 10^{-6}, & r_5 &= 7.83\times 10^{-3}, & r_6 &= 3.38\times
10^{-3}, & r_7 &= -9.89\times 10^{-6}, \\
r_8 &= 0.0939, & r_9 &= 0.204, & r_{10} &= -0.103.
\end{aligned}
\end{equation}
This agrees with the exact result within maximal deviations of $2.7\times
10^{-5}$ for 10 GeV $\leq
M_H \leq$ 1~TeV and the other input parameters in their $2\sigma$
ranges, corresponding to an error of 0.4 MeV for $\MW$ and $8\times 10^{-6}$ for
$\sineff$.
For the top quark mass and the $W$-boson mass varying in their $4\sigma$
ranges, the formula eq.~\eqref{eq:formdr} is
accurate to $4.3\times 10^{-5}$, corresponding to an error of 0.65 MeV for $\MW$
and $12.5\times 10^{-6}$ for $\sineff$.

\subsection{Results for other fermion flavors}
\label{sec:other}

The results presented in the previous sections and in
Refs.~\cite{sineff,sinbos} give the effective weak mixing angle \SinEff defined
for the leptonic $Zl^+l^-$ vertex. For the $Zf\bar{f}$ vertex with other light
flavors $f=\nu,u,d$ in the final state, there are small but non-zero
differences with respect to the leptonic effective weak mixing angle. 
In this section, results are given for \SinEfff\ for different final state
fermions except b-quarks. 
For the $b\bar{b}$ final state, the two-loop electroweak corrections are still
missing, since they involve new topologies with additional top-quark
propagators.

Since the numerical effect of the fermionic electroweak two-loop
corrections is much larger than the corresponding bosonic contributions, only
the fermionic \Oaa\ diagrams are taken into account. As before, the complete
one-loop corrections and the (flavor independent) contributions of order
${\cal O}(\al\alps)$, ${\cal O}(\al\alps^2)$, ${\cal O}(\al^2\alps\mt^4)$ and
${\cal O}(\al^3 \mt^6)$ are also included.

As before, the numerical results are expressed through the parametrization
in eq.~\eqref{eq:formula}, which reproduces the exact calculation with maximal
deviations of $4.5\times10^{-6}$, when the input parameters stay within
their $2\sigma$
ranges and the Higgs boson mass in the range 10 GeV $\leq M_H \leq$ 1 TeV.
The values of the coefficients for the various final state flavors are listed in
Tab.~\ref{tab:flav}.
\begin{table}[tb]
\begin{center}
\begin{tabular}{l@{$\quad$}llll}
\hline
$f$ & $e,\mu,\tau$ & $\nu_{e,\mu,\tau}$ & $u,c$ & $d,s$ \\
\hline
$s_0$ & 0.2312527 & 0.2308772 & 0.2311395 & 0.2310286 \\
$d_1$ [$10^{-4}$] & \phantom{$-$}4.729 & \phantom{$-$}4.713 & \phantom{$-$}4.726 & \phantom{$-$}4.720 \\
$d_2$ [$10^{-5}$] & \phantom{$-$}2.07 & \phantom{$-$}2.05 & \phantom{$-$}2.07 & \phantom{$-$}2.06 \\
$d_3$ [$10^{-6}$] & \phantom{$-$}3.85 & \phantom{$-$}3.85 & \phantom{$-$}3.85 & \phantom{$-$}3.85 \\
$d_4$ [$10^{-6}$] & $-$1.85 & $-$1.85 & $-$1.85 & $-$1.85  \\
$d_5$ [$10^{-2}$] & \phantom{$-$}2.07 & \phantom{$-$}2.06 & \phantom{$-$}2.07 & \phantom{$-$}2.07 \\
$d_6$ [$10^{-3}$] & $-$2.851 & $-$2.850 & $-$2.853 & $-$2.848 \\
$d_7$ [$10^{-4}$] & \phantom{$-$}1.82 & \phantom{$-$}1.82 & \phantom{$-$}1.83 & \phantom{$-$}1.81 \\
$d_8$ [$10^{-6}$] & $-$9.74 & $-$9.71 & $-$9.73 & $-$9.73\\
$d_9$ [$10^{-4}$] & \phantom{$-$}3.98 & \phantom{$-$}3.96 & \phantom{$-$}3.98 & \phantom{$-$}3.97 \\
$d_{10}\!\!$ [$10^{-1}$] & $-$6.55 & $-$6.54 & $-$6.55 & $-$6.55 \\
\hline
\end{tabular}
\end{center}
\vspace{-1em}
\caption{Coefficient of the fitting formulae eq.~\eqref{eq:formula} for
different final states $f\bar{f}$.
\label{tab:flav}}
\end{table}

\subsection{Implementation into global Standard Model fits}

The fermionic two-loop corrections and some higher-order contributions as
listed in eq.~\eqref{eq:swfull} are implemented in the current version 6.42 of
the program {\sc Zfitter} \cite{zfitter,zfitternew}, which is widely used for
global fits of the Standard Model to electroweak precision data \cite{lepewwg}.
Due to the complexity of the two-loop computation, the implementation of the
exact result was not possible, so that instead the numerical fitting formula
eq.~\eqref{eq:formula} was included in the code. More details can be found in
Ref.~\cite{zfitternew}.

The fitting formula has been incorporated exactly only for the leptonic
effective weak mixing angle \SinEff, i.e.\ for the $Zl^+l^-$ vertex. Results
for other light flavors $f=u,d,c,s,\nu$ in the final state are implemented in
an approximate way, which reproduces the complete results of
section~\ref{sec:other} within an error of about $10^{-5}$ for $f=u,d,c,s$ and
$2 \times 10^{-5}$ for $f=\nu$.

For the $b\bar{b}$ final state, no two-loop electroweak corrections beyond the
leading $\mt^4$ are included in {\sc Zfitter 6.42}. They shall become available
in a future version. However, the current version 6.42 was adjusted with respect
to previous version to include complete two-loop corrections in the initial
state vertex for the process $e^+e^- \to (Z) \to b\bar{b}$, see
Refs.~\cite{zfitternew,zbb} for details.


\section{Conclusion}
\label{sc:concl}

In this paper, the evaluation of the complete two-loop contributions
to the effective weak mixing angle has been described, expatiating the
computational methods and the quantitative implications of the new result.

It was shown how the effective weak mixing angle can be defined at
next-to-next-to-leading order through the vector and axial-vector couplings of
the $Z$-boson. The computation of the vertex loop diagrams using two independent
techniques for the fermionic part and a combination of several computational
methods for the bosonic part was elucidated in detail.

Numerical results for the effective weak mixing angle for different final state
flavors were given in terms of accurate numerical parameterizations, which are
valid for Higgs masses up to 1 TeV. The new result has been compared in detail
with a previous result obtained by an expansion in powers of
$\mt$ up to next-to-leading order.

Furthermore, the remaining theoretical uncertainties due to unknown higher
orders were analyzed and an overall uncertainty of the effective leptonic weak
mixing angle \SinEff\ of $4.7 \times 10^{-5}$ was estimated.

Electroweak precision data allows very precise tests of the Standard Model at
the quantum level and puts the strongest constraints on the Higgs boson
mass and new physics. With the completion of the electroweak two-loop
corrections, the accuracy of the electroweak precision test was significantly
enhanced, with theoretical uncertainties now under much better control.

\bigskip

\vspace{- .3 cm}
\section*{Acknowledgements}
We are grateful to G.~Weiglein and K.~M\"onig for useful discussions and
communications. We thank T.~Riemann for helping to update the new version of 
{\sc Zfitter}.

    The work of M.~A. was supported by the BMBF grant No. 05 HT4GUA/4 and
by the DFG grant No. SFB 676.
    The work of M.~C. was supported by the Sofja
    Kovalevskaja Award of the Alexander von Humboldt Foundation
    sponsored by the German Federal Ministry of Education and
    Research. A.~F. is supported by the Schweizer Nationalfonds.


\end{document}